\documentclass[12pt]{article}
\usepackage{amsfonts}
\usepackage{latexsym}
\usepackage{amsmath}
\usepackage{amssymb}
\usepackage{amssymb}
\usepackage{mathrsfs}

\newcommand{\newsection}{    
\setcounter{equation}{0}\section}
\def\appendix#1{\addtocounter{section}{1}\setcounter{equation}{0}
\renewcommand{\thesection}{\Alph{section}}
\section*{Appendix \thesection\protect\indent \parbox[t]{11.15cm}{#1}}
\addcontentsline{toc}{section}{Appendix \thesection\ \ \ #1}}

\newcommand{\be}{\begin{eqnarray}}
\newcommand{\ee}{\end{eqnarray}}
\newcommand{\bea}{\begin{eqnarray}}
\newcommand{\eea}{\end{eqnarray}}
\newcommand{\ba}{\begin{array}}
\newcommand{\ea}{\end{array}}

\newenvironment{frcseries}{\fontfamily{frc}\selectfont}{}

\def \la {\label}

\def\e{\epsilon}

\def\bbe{{\bf{e}}}
\font\mybb=msbm10 at 11pt

\def\bb#1{\hbox{\mybb#1}}

\def\bR {\bb{R}}

\def\bC {\bb{C}}

\def\tn {{\tilde{\nabla}}}
\def\bI {\bb{I}}
\def\bS {\bb{S}}

\def\tn {{\tilde{\nabla}}}

\begin{document}
\begin{titlepage}
\begin{center}
\vspace*{-1.0cm}
\vspace*{-2cm} 
\begin{flushright}
{\tt KCL-MTH-18-05}\\
\end{flushright}

\vspace{2.0cm} {\Large \bf  Symmetry enhancement of extremal horizons in $D=5$ supergravity} \\[.2cm]

\vspace{1.5cm}
 {\large  U.~ Kayani}

\vspace{0.5cm}

\vspace{0.5cm}
Department of Mathematics\\
King's College London\\
Strand\\
London WC2R 2LS, UK\\

\vspace{0.5cm}

\end{center}
\vskip 1.5 cm
\begin{abstract}

\end{abstract} We consider the near-horizon geometry of supersymmetric extremal black holes
in un-gauged and gauged 5-dimensional supergravity, coupled to abelian vector multiplets. By analyzing the
global properties of the Killing spinors, we prove that the near-horizon geometries
undergo a supersymmetry enhancement. This follows from a set of generalized Lichnerowicz-type
theorems we establish, together with an index theory argument. As a consequence, these solutions 
always admit a $\mathfrak{sl}(2,\mathbb{R})$ symmetry group.

\end{titlepage}



\setcounter{section}{0}
\setcounter{subsection}{0}
\setcounter{equation}{0}

\newsection{Introduction}

The enhancement of supersymmetry near to brane and black hole horizons has been known
for some time. In the context of branes, many solutions are known which exhibit supersymmetry
enhancement near to the brane. For example, the geometry of D3-branes  doubles its supersymmetry to become the maximally supersymmetric $AdS_5 \times S^5$
solution \cite{gwgpkt, Schwarz:1983qr}. This phenomenon played a crucial role in the early development of the
ADS/CFT correspondence \cite{maldacena}. Black hole solutions are also known to exhibit
supersymmetry enhancement; for example in the case of the five-dimensional BMPV black hole \cite{BMPV1, Chamseddine:1996pi, Chamseddine:1999qs}. 

The black hole horizon topology is important in establishing black hole uniqueness theorems. In $D=4$ these imply that the Einstein equations admit a unique class of asymptotically flat black hole solutions, parametrized by $(M, Q, J)$. A key step is to establish the horizon topology theorem, which proves that the event horizon of a stationary black hole must have $S^2$ topology \cite{Htopology1}. This relies on the Gauss-Bonnet theorem applied to the 2-manifold spatial horizon section, and therefore does not generalize to higher dimensions. Indeed, the first example of how the classical uniqueness theorems 
break down in higher dimensions is given by the five-dimensional black ring solution
\cite{Emparan:2001wn, BR1}. There exist black ring solutions with the same asymptotic
conserved charges as BMPV black holes, but with a different horizon topology.  Even more exotic solutions in five dimensions are now known to exist,
such as the solutions obtained in \cite{Horowitz:2017fyg}, describing asymptotically flat black holes which possess a non-trivial topological structure outside the event horizon, but whose
near-horizon geometry is the same as that of the BMPV solution.

Another important observation in the study of black holes is the attractor mechanism.
This states that the entropy is obtained by extremizing an entropy function which depends only on
the near-horizon parameters and conserved charges, and if this admits a unique extremum then the entropy is independent of the asymptotic values of the moduli.
In the case of 4-dimensional solutions the analysis of \cite{astef} implies that if
the solution admits $SO(2,1) \times U(1)$ symmetry, and the horizon has 
spherical topology, then such a mechanism holds. 
In $D=4, 5$ it is an observation that all known asymptotically flat black hole solutions exhibit attractor mechanism behaviour.
However, in higher dimensions, it is unclear if an attractor mechanism holds. In particular, a generalization of the analysis
of \cite{astef} to higher dimensions would require the existence of
a $SO(2,1) \times U(1)^2$ symmetry, as well as an understanding of the horizon topology. Near horizon geometries of asymptotically $AdS_5$ 
supersymmetric black holes admitting a $SO(2,1) \times U(1)^2$ symmetry have been 
classified in \cite{Kunduri:2006ek, Kunduri:2007qy}. It remains to be determined
if all supersymmetric near-horizon geometries fall into this class.

Further recent interest in the geometry of black hole horizons has arisen in the context of
the BMS-type symmetries associated with black holes, following 
\cite{Hawking:2016msc, Hawking:2016sgy, Averin:2016ybl, Donnay:2015abr}.
In particular, the analysis of the asymptotic symmetry group of Killing horizons was undertaken in
\cite{Akhmedov:2017ftb}. In that case, an infinite dimensional symmetry group is obtained,
analogous to the BMS symmetry group of asymptotically flat solutions. 

In this paper we shall investigate the mechanism by which supersymmetry is enhanced
for supersymmetric extremal black hole near-horizon geometries in both gauged and ungauged $N=2$, $D=5$ supergravity. We will assume that the black hole event horizon is a Killing horizon. Rigidity theorems
have been constructed which imply that the black hole horizon is Killing
for both non-extremal and extremal black holes, under certain assumptions, have been constructed, e.g. \cite{rigidity1, gnull, axi1, axi2}. The assumption that the event horizon is Killing
enables the introduction of Gaussian Null co-ordinates \cite{isen, gnull} in a neighbourhood
of the horizon. The analysis of the near-horizon geometry is significantly simpler than that of the
full black hole solution, as the near-horizon limit reduces the system to a set of
equations on a co-dimension 2 surface, ${\cal{S}}$, which is the spatial section of
the event horizon.

The proof that we give in this paper for (super)symmetry enhancement relies on establishing Lichnerowicz-type theorems and an index theory argument. A similar proof has been given for supergravity horizons in $D=11, D=10$ for IIA, Roman's Massive IIA and IIB, $D=5$ minimal gauged and $D=4$ gauged \cite{11index, iiaindex, miiaindex, iibindex, 5dindex, 4dindex}. 
We shall also prove that the near-horizon geometries admit a $\mathfrak{sl}(2,\mathbb{R})$
symmetry algebra. In general we find that the orbits of the generators of $\mathfrak{sl}(2,\mathbb{R})$ are
3-dimensional, though in some special cases they are 2-dimensional. In these special cases,
the geometry is a warped product $AdS_2 \times_w {\cal{S}}$.
The properties of $AdS_2$ and their relationship to black hole entropy
have been examined in \cite{strominger, sen}.
Our result, together with those of our previous calculations, implies 
that the $\mathfrak{sl}(2,\mathbb{R})$ symmetry is a universal property of supersymmetric black holes.

Previous work has also been done on the classification of near-horizon geometries for
five dimensional ungauged supergravity in \cite{gutbh, reall}. However there an additional assumption was made on assuming the vector bilinear matching condition i.e the black hole Killing horizon associated with a Killing vector field is identified as a Killing spinor bilinear. We do not make this assumption here, and we prove the results on (super)symmetry enhancement in full generality. The only assumptions we make in the paper are that all the fields are smooth (or at least $C^2$ differentiable) and the spatial horizon section ${\cal S}$ is compact, connected and without boundary. These assumptions are made in order that  various global techniques can be applied to the analysis.

The content in this paper is organised in the following way. In section 2, we state the key properties for $D=5, N=2$ gauged supergravity, coupled to an arbitrary number of vector multiplets. We give the bosonic part of the action, the field equations and the fermionic supersymmetry variations (the vanishing of which are the KSEs). In section 3, we solve the KSEs by appropriately decomposing the gauge fields and integrating along two lightcone directions. and  we identify the independent KSEs. In section 4, we establish a generalized Lichnerowicz-type theorem in order to show the, on spatial cross-sections of the event horizon, the zero modes certain Dirac operators ${\mathscr D}^{(\pm)}$ are in a 1-1 correspondence with the Killing spinors. In section 5, we prove the supersymmetry enhancement, and we analyse the relationship between positive and negative lightcone chirality spinors which gives rise to the doubling of the supersymmetry. We also prove that horizons with non-trivial fluxes admit an ${\mathfrak{sl}}(2, \mathbb{R})$ symmetry subalgebra. 

In appendix A, we state the supersymmetry conventions. In appendix B, we state the spin connection and the Ricci curvature tensor. In appendix C, we state the independent horizon Bianchi identities and field equations. In section D, we state the independent horizon Bianchi identities and field equations for the gauge decomposition given in section 3. 
In Appendix E we present some details of the calculations used to find the minimal set of independent KSEs on the spatial horizon section.
In appendix F, we prove the scalar orthogonality condition, which is used to simplify the KSEs and field equations in section 2.

\newsection{$D=5, N=2$ Gauged Supergravity}

In this section, we briefly summarize some of the key properties of $D=5$, $N=2$ gauged supergravity,
coupled to $k$ vector multiplets. The bosonic part of the action is associated with a particular hypersurface $N$ of $\mathbb{R}^k$ defined by
\bea
V(X) = \frac{1}{6}C_{I J K}X^I X^J X^K = 1
\eea
where the fields $\{ X^I = X^I(\phi) \, , I=0,\dots,k-1 \}$ are standard coordinates on $\mathbb{R}^k$; and where $X_I$, the dual coordinate is defined by,
\bea
X_{I} = \frac{1}{6}C_{I J K}X^J X^K
\eea
and $C_{I J K}$ are constants which are symmetric in $IJK$. This allows us to express the hypersurface equation $V=1$ as $X^I X_I = 1$ and one can deduce that
\bea
\label{hyre}
\partial_a X_I &=& \frac{1}{3}C_{I J K}\partial_a X^J X^K 
\nonumber \\
X^I \partial_a X_I &=& X_I \partial_a X^I = 0 \ .
\eea
The bosonic part of the supergravity action is given by,
\bea
S_{bos} &=&\int d^5 x \sqrt{-g} \bigg(R - \frac{1}{2}Q_{I J}(\phi)F^{I}{}_{\mu \nu} F^{J \mu \nu} - h_{a b}(\phi)\partial_{\mu} \phi^a \partial^{\mu} \phi^b  + 2 \chi^2 U \bigg)
\nonumber \\
&&+ \frac{1}{24}e^{\mu \nu \rho \sigma \tau}C_{I J K}F^{I}{}_{\mu \nu}F^{J}{}_{\rho \sigma}A^{K}{}_{\tau} 
\eea
where $F^I = dA^{I}$, $I, J, K = 0, \dots, k-1$ are the 2-form Maxwell field strengths,
$\phi^a$ are scalars, $\mu, \nu, \rho, \sigma = 0, \dots, 4$, and $g$ is the metric of the five-dimensional spacetime, and $U$ is the scalar potential which can be expressed as,
\bea
U = 9V_I V_J\bigg(X^I X^J - \frac{1}{2}Q^{I J}\bigg)
\eea
where $V_{I}$ are constants. The gauge coupling $Q_{I J}$, and the metric $h_{a b}$ on $N$ are given by,
\bea
\label{gcmetric}
Q_{I J} = -\frac{1}{2}\frac{\partial}{\partial X^I}\frac{\partial}{\partial X^J}(\ln {V})|_{{V} =1} = -\frac{1}{2}C_{I J K}X^K + \frac{9}{2}X_{I}X_J \ 
\eea
and
\bea
\label{pbmetric}
h_{a b} = Q_{I J}\frac{\partial X^I}{\partial \phi^a}\frac{\partial X^J}{\partial \phi^b}|_{{V} =1}
\eea
where $\{ \phi^a \, , a=1,\dots,k-1 \}$ are local coordinates of $N$. 
We shall assume that the gauge coupling $Q_{IJ}$ is positive definite, and also
that the scalar potential is non-negative, $U \geq 0$.

In the case of the $STU$ model, which has $C_{123}=1$, and $X^1 X^2 X^3=1$, the non-vanishing components of
the gauge coupling are given by
\bea
Q_{11}={1 \over 2(X^1)^2}, \quad Q_{22}={1 \over 2(X^2)^2}, 
\quad Q_{33}={1 \over 2(X^3)^2}
\eea
with scalar potential
\bea
U= 18 \bigg({V_1 V_2 \over X^3}+{V_1 V_3 \over X^2}+{V_2 V_3 \over X^1}\bigg) \ .
\eea
When considering near-horizon solutions, conditions which are sufficient
to ensure that $U \geq 0$ are
that $V_I \geq 0$ for $I=1,2,3$, and also that there exists a point on
the horizon section at which $X^I >0$ for $I=1,2,3$. As we shall assume
that the scalars are smooth functions on (and outside of) the horizon, this implies that $X^I>0$ everywhere on the horizon.

In addition, the
following relations also hold:
\bea
X_I &=& \frac{2}{3}Q_{I J}X^J
\nonumber \\
\partial_a X_I &=& - \frac{2}{3}Q_{I J}\partial_a X^J \ .
\eea
The Einstein equation is given by
\bea
\label{eins}
R_{\mu\nu}- Q_{I J}\left(  F^{I}{}_{\mu\lambda}F^{J}{}_{\nu}{}^{\lambda}
+\nabla_{\mu}X^{I}\nabla_{\nu}X^{J}-\frac{1}{6}g_{\mu\nu}F^{I}{}_{\rho\sigma
}F^{J\rho\sigma}\right) + \frac{2}{3}\chi^2 U g_{\mu \nu} = 0 \ .
\eea
The Maxwell gauge equations for $A^I$ are given by
\bea
\label{maxwell}
d(Q_{I J}\star_5 F^J) = \frac{1}{4}C_{I J K}F^J \wedge F^K \ ,
\eea
or equivalently, in components:
\bea
\label{maxeq}
\nabla_{\mu}(Q_{I J}F^{J \mu \nu}) = -\frac{1}{16}C_{I J K}e^{\nu \mu \rho \sigma \tau}F^J{}_{\mu \rho}F^K{}_{\sigma \tau}
\eea
where $e^{\mu \nu \rho \sigma \kappa} = \sqrt{-g}\epsilon^{\mu \nu \rho \sigma \kappa}$. The scalar field equations for $\phi^a$ are
\bea
\label{scalar}
&&\bigg[\nabla^{\mu}\nabla_{\mu}X_{I} + \bigg(-\frac{1}{6}C_{M N I} + X_M X^P C_{N P I} \bigg)\bigg(\frac{1}{2}F^M{}_{\mu \nu}F^{N \mu \nu} + \nabla_{\mu}X^M \nabla^{\mu} X^N \bigg) 
\nonumber \\
&&+ \frac{3}{2}\chi^2 C_{I J K}Q^{M J}Q^{N K}V_{M}V_{N}\bigg] \partial_a X^I = 0 \ .
\eea

We remark that if $L_{I}\partial_{a}{X^{I}} = 0$ 
for all $a=1, \dots , k-1$, then $L_{I} = f X_{I}$ where $f = X^{J}L_{J}$.
This result is established in Appendix E.
Using this, the scalar field equation can be rewritten as
\bea
\label{scalareq1}
&&\nabla^{\mu}\nabla_{\mu}{X_{I}} + \nabla_{\mu}{X^{M}} \nabla^{\mu}{X^{N}} \left( \frac{1}{2}C_{M N K} X_{I} X^{K} - \frac{1}{6}C_{I M N}\right) 
\nonumber \\
&+& \frac{1}{2}F^{M}{}_{\mu \nu} F^{N \mu \nu} \left(C_{I N P} X_{M} X^{P} - \frac{1}{6}C_{I M N}-6X_{I} X_{M} X_{N}+\frac{1}{6}C_{M N J} X_{I} X^{J}\right) 
\nonumber \\
&+& 3 \chi^2 V_{M} V_{N}\bigg(\frac{1}{2}C_{I J K}Q^{M J}Q^{N K} + X_{I}(Q^{M N} - 2 X^{M}X^{N})\bigg) = 0 \ .
\eea

\newsection{Evaluation of Killing Spinor Equations}
The KSEs are defined on a purely bosonic background, and are given as the vanishing of the supersymmetry transformations of the fermions at lowest order in fermions. The number of linearly independent Killing spinors determines how much supersymmetry is realised for a given solution. The KSEs can be expressed as,
\bea
{\cal D}_\mu\e&\equiv& \nabla_\mu \epsilon + \frac{i}{8}X_I\bigg(\Gamma_{\mu}{}^{\nu \rho} - 4\delta_{\mu}{}^{\nu}\Gamma^{\rho}\bigg)F^I{}_{\nu \rho} \epsilon + \bigg(- \frac{3i}{2}\chi V_{I}A^{I}{}_{\mu} + \frac{1}{2}\chi V_I X^{I}\Gamma_\mu\bigg)\epsilon  = 0
\nonumber \\
\label{Gkseo} \\
\label{Akseo}
{\cal A}^{I}\e &\equiv& \bigg[\bigg(\delta^{J}{}_{I} - X^{I}X_{J}\bigg)F^{J}{}_{\mu \nu}\Gamma^{\mu \nu} + 2i\Gamma^{\mu}\partial_{\mu}X^I - 6i\chi \bigg(Q^{I J} - \frac{2}{3}X^{I}X^{J}\bigg)V_J\bigg]\e = 0 \ .
\nonumber \\
\eea
On decomposing $F^I$ as 
\bea
\label{decom}
F^I = FX^I + G^I
\eea
where
\bea
X_I F^I = F, \qquad X_I G^I = 0 \ .
\eea 
the KSEs can then be rewritten in terms of $F$ and $G^I$ as
\bea
\hspace{-1cm}{\cal D}_\mu\e&\equiv& \nabla_\mu \epsilon + \frac{i}{8}\bigg(\Gamma_{\mu}{}^{\nu \rho} - 4\delta_{\mu}{}^{\nu}\Gamma^{\rho} \bigg)F_{\nu \rho} \epsilon + \bigg(- \frac{3i}{2}\chi V_{I}A^{I}{}_{\mu} + \frac{1}{2}\chi V_I X^{I}\Gamma_\mu\bigg)\epsilon = 0 \ , \label{Gkse} 
\eea
and
\bea
{\cal A}^{I} \e &\equiv& \bigg[G^{I}{}_{\mu \nu}\Gamma^{\mu \nu} + 2i\Gamma^{\mu}\partial_{\mu}X^I - 6i\chi \bigg(Q^{I J} - \frac{2}{3}X^{I}X^{J}\bigg)V_J \bigg] \e = 0  \ .
\label{Akse}
\eea

\subsection{Near-horizon Data}

In order to study near-horizon geometries we need to introduce a coordinate system which is regular and adapted to the horizon. We will consider a five-dimensional stationary black hole metric,
for which the horizon is a Killing horizon, and the metric is regular at the horizon. A set of Gaussian Null coordinates \cite{isen, gnull} $\{u, r, y^{I}\}$
will be used to describe the metric, where $r$ denotes the radial distance away from the event horizon which is located at $r=0$ and $y^I,~ I=1, \dots, 3$ are local co-ordinates on ${\cal S}$. The metric components have no dependence on $u$, and the timelike isometry ${\partial \over \partial u}$ is null on the horizon at $r=0$. The black hole metric in a patch containing the horizon is given by
\bea
\label{gnulmet}
ds^2 = 2du dr + 2r h_I(r, y) du dy^I - r f(r, y) du^2 + ds_{\cal S}^2 \ .
\eea
The spatial horizon section ${\cal S}$ is given by $u=const,~ r=0$ with the metric 
\bea
ds_{\cal S}^2  = \gamma_{I J}(r, y)dy^I dy^J \ .
\eea
We assume that ${\cal{S}}$ is compact, connected and without boundary. The
1-form $h$, scalar $\Delta$ and metric $\gamma$ are functions of $r$ and $y^{I}$; they  are analytic in $r$ and regular at the horizon. The surface gravity associated with the Killing horizon is given by $\kappa = \frac{1}{2}f(y,0)$. The near-horizon limit is a particular decoupling limit defined by
\bea
\label{nhl}
r \rightarrow \epsilon r,~ u \rightarrow \epsilon^{-1} u,~ y^{I} \rightarrow y^{I}, \qquad {\rm and} \qquad \epsilon \rightarrow 0 \ .
\eea
This limit is only defined when $f(y,0) = 0$, which implies that the surface gravity vanishes, $\kappa = 0$. Hence the near horizon geometry is only well defined for extreme black holes,
and we shall consider only extremal black holes here. After taking the limit (\ref{nhl}) we obtain, 
\bea
\label{nhm}
ds_{NH}^2 = 2du dr + 2r h_I(y) du dy^I - r^2 \Delta(y) du^2  + \gamma_{I J}(y)dy^I dy^J \ .
\eea
In particular, the form of the metric remains unchanged from ({\ref{gnulmet}}), however
the 1-form $h$, scalar $\Delta$ and metric $\gamma$ on ${\cal{S}}$ no longer have any radial dependence
{\footnote{The near-horizon metric (\ref{nhm}) also has a new scale symmetry, $r \rightarrow \lambda r,~ u \rightarrow \lambda^{-1}u$ generated by the Killing vector $L=u\partial_{u} - r\partial_{r}$. This, together with the Killing vector $V=\partial_u$ satisfy the algebra $[V, L] = V$ and they form a 2-dimensional non-abelian symmetry group ${\cal{G}}_2$. We shall show that this further enhances into a larger symmetry algebra, which will include a $\mathfrak{sl}(2,\mathbb{R})$ subalgebra.}}. For $N=2$, $D=5$ supergravity, in addition to the metric, there are also gauge field strengths and
scalars. We will assume that these are also analytic in $r$ and regular at the horizon, 
and that there is also a consistent near-horizon limit for these matter fields:
\bea
A^{I} &=& -r \alpha^I \bbe^+ + {\tilde{A}}^{I}
\nonumber \\
F^I &=& \bbe^+ \wedge \bbe^- \alpha^I + r \bbe^+ \wedge \beta^I + {\tilde F}^I~,
\la{hormetr}
\eea
where $F^{I} = dA^{I}$ and we have introduced the frame
\be
\label{basis1}
\bbe^+ = du, \qquad \bbe^- = dr + rh -{1 \over 2} r^2 \Delta du, \qquad \bbe^i = e^i{}_I dy^I~,
\ee
in which the metric is
\bea
\label{nhf}
ds^2 &=&2 \bbe^+ \bbe^- + \delta_{ij} \bbe^i \bbe^j \ .
\eea
We can also express the near horizon fields $F$ and $G^I$ in this frame as
\bea
\label{nhfs}
F &=& \bbe^+ \wedge \bbe^- \alpha + r \bbe^+ \wedge \beta + {\tilde F}
\nonumber \\
G^I &=& \bbe^+ \wedge \bbe^- L^I + r \bbe^+ \wedge M^I + {\tilde G}^I
\eea
where $X_I L^I = X_I M^I = X_I \tilde{G}^I = 0$ and we set $\alpha = X_I \alpha^I$, $\tilde{F} = X_I \tilde{F}^I$ and $\beta = X_I \beta^I$.

\subsection{Solving the KSEs along the Lightcone}

For supersymmetric near-horizon horizons we assume there exists an $\epsilon \neq 0$ which is a solution to the KSEs. In this section, we will determine the neccessary conditions on the Killing spinor. To do this we first integrate along the two lightcone directions i.e.~we integrate the KSEs along the $u$ and $r$ coordinates. To do this, we decompose $\epsilon$ as
\bea
\e=\e_++\e_-~,
\label{ksp1}
\eea
where $\Gamma_\pm\epsilon_\pm=0$, and find that
\bea\label{lightconesol}
\e_+=\phi_+(u,y)~,~~~\e_-=\phi_-+r \Gamma_-\Theta_+ \phi_+~,
\eea
and
\bea
\phi_-=\eta_-~,~~~\phi_+=\eta_++ u \Gamma_+ \Theta_-\eta_-~,
\eea
where
\bea
\label{thetapm}
\Theta_\pm &=& {1\over4} h_i\Gamma^i - \frac{i}{8}(\tilde{F}_{j k}\Gamma^{j k}  \pm 4\alpha) - \frac{1}{2}\chi V_I X^I 
\eea
and $\eta_\pm$ depend only on the coordinates of the spatial horizon section ${\cal S}$.
Substituting the solution (\ref{lightconesol}) of the KSEs along the light cone directions back into the gravitino KSE (\ref{Gkse}), and appropriately expanding in the $r$ and $u$ coordinates, we find that
for  the $\mu = \pm$ components, one obtains  the additional conditions
\bea
\label{int1}
& &\bigg({1\over2}\Delta - {1\over8}(dh)_{ij}\Gamma^{ij} -\frac{i}{4}\beta_i \Gamma^i + \frac{3i}{2}\chi V_{I}\alpha^I \bigg)\phi_+ 
\nonumber \\
&+& 2\bigg({1\over4} h_i\Gamma^i - \frac{i}{8}(-\tilde{F}_{j k}\Gamma^{j k} + 4\alpha) + \frac{1}{2} \chi V_{I}X^{I}\bigg)\tau_+ = 0~,
\eea
\bea
\label{int2}
\bigg(\frac{1}{4}\Delta h_i \Gamma^{i} - \frac{1}{4}\partial_{i}\Delta \Gamma^{i}\bigg)\phi_+ + \bigg(-\frac{1}{8}(dh)_{ij}\Gamma^{ij} +\frac{3i}{4}\beta{}_i\Gamma^i  + \frac{3i}{2}\chi V_{I}\alpha^I\bigg) \tau_+ = 0~,
\eea
\bea
\label{int3}
& &\bigg(-\frac{1}{2}\Delta - \frac{1}{8}(dh)_{ij}\Gamma^{ij} -\frac{3i}{4}\beta{}_i \Gamma^i + \frac{3i}{2}\chi V_{I}\alpha^I 
\nonumber \\
&+& 2\big(-{1\over4} h_i\Gamma^i - \frac{i}{8}(\tilde{F}_{j k}\Gamma^{j k} + 4\alpha) - \frac{1}{2}\chi V_{I}X^{I}\big) \Theta_{-} \bigg)\phi_{-} = 0 \ .
\eea
Similarly the $\mu=i$ component of the gravitino KSEs gives
\bea
\label{int4}
\tilde{\nabla}_i \phi_\pm + \bigg( \mp \frac{1}{4}h_i  \mp \frac{i}{4}\alpha \Gamma_i  + \frac{i}{8}\tilde{F}_{j k}\Gamma_i{}^{j k} - \frac{i}{2}\tilde{F}_{i j}\Gamma^j - \frac{3i}{2}\chi V_{I}\tilde{A}^{I}{}_{i} + \frac{1}{2}\chi V_I X^{I}\Gamma_i \bigg) \phi_\pm=0~,~~~
\eea
and
\bea
\label{int5}
&&\tilde \nabla_i \tau_{+} + \bigg( -\frac{3}{4}h_i - \frac{i}{4}\alpha\Gamma_i - \frac{i}{8}\tilde{F}_{j k}\Gamma_i{}^{j k} + \frac{i}{2}\tilde{F}_{i j}\Gamma^j - \frac{3i}{2}\chi V_{I}\tilde{A}^{I}{}_{i} - \frac{1}{2}\chi V_I X^{I}\Gamma_i\bigg )\tau_{+}
\nonumber \\
&&+ \bigg(-\frac{1}{4}(dh)_{ij}\Gamma^{j} - \frac{i}{4}\beta_j \Gamma_i{}^j + \frac{i}{2}\beta_i   \bigg)\phi_{+} = 0~,
\eea
where we have set
\bea
\label{int6}
\tau_{+} = \Theta_{+}\phi_{+} \ .
\eea
Similarly, substituting the solution of the KSEs (\ref{lightconesol})  into the algebraic KSE (\ref{Akse}) and expanding appropriately in the $u$ and $r$ coordinates, we find
\bea
\label{int7}
\bigg[\tilde{G}^I{}_{i j}\Gamma^{i j} \mp 2 L^I + 2i\tilde{\nabla}_i X^I \Gamma^i - 6i\chi \bigg(Q^{I J} - \frac{2}{3}X^{I}X^{J}\bigg)V_J\bigg]\phi_\pm = 0  \ ,
\eea
\be
\label{int8}
\hspace{-1cm}\bigg[\tilde{G}^I{}_{i j}\Gamma^{i j} + 2 L^I - 2i\tilde{\nabla}_i X^I \Gamma^i - 6i\chi \bigg(Q^{I J} - \frac{2}{3}X^{I}X^{J}\bigg)V_J \bigg]\tau_{+} + 2 M^I{}_i\Gamma^i \phi_{+}=0~.
\eea
In the next section, we will demonstrate that many of the above conditions are redundant as they are implied by the independent KSEs\footnote{These are given by the naive restriction of the KSEs on ${\cal S}.$} (\ref{covr}), upon using the field equations and Bianchi identities.

\subsection{The Independent KSEs on $\cal{S}$}

The integrability conditions of the KSEs in any supergravity theory are known to imply some of the Bianchi identities and field equations. Also, the KSEs are first order differential equations which are usually easier to solve than the field equations which are second order. As a result, the standard approach to find solutions is to first solve all the KSEs and then impose the remaining independent components of the field equations and Bianchi identities as required.
We will take a different approach here because of the difficulty of solving the KSEs and the algebraic conditions which include the $\tau_+$ spinor given in (\ref{int6}). Furthermore,  we are particularly interested
in the minimal set of conditions required for supersymmetry, in order to systematically analyse the necessary and
sufficient conditions for supersymmetry enhancement. 

In particular, the conditions  (\ref{int1}), (\ref{int2}), (\ref{int5}), and (\ref{int8}) which contain $\tau_+$ are implied from those containing $\phi_+$, along with some of the field equations and Bianchi identities. Furthermore, (\ref{int3}) and the terms linear in $u$ in (\ref{int4}) and (\ref{int7}) from the $+$ component are implied by the field equations, Bianchi identities and the $-$ component of (\ref{int4}) and (\ref{int7}).
Details of the calculations used to show this are presented in Appendix E.

On taking this into account, it follows that, on making use of the field equations and Bianchi identities, the independent KSEs are
\bea
\label{covr}
\nabla^{(\pm)}_{i} \eta_{\pm} = 0, \qquad {\cal A}^{I, (\pm)}\eta_{\pm} = 0
\eea
where
\bea
\nabla^{(\pm)}_{i} = \tilde{\nabla}_{i} + \Psi^{(\pm)}_{i}
\eea
with
\bea
\label{alg1pm}
\Psi^{(\pm)}_{i} &=& \mp \frac{1}{4}h_i  \mp \frac{i}{4}\alpha \Gamma_i  + \frac{i}{8}\tilde{F}_{j k}\Gamma_i{}^{j k} - \frac{i}{2}\tilde{F}_{i j}\Gamma^j - \frac{3i}{2}\chi V_{I}\tilde{A}^{I}{}_{i} + \frac{1}{2}\chi V_I X^{I}\Gamma_i \ ,
\eea
and
\bea
\label{alg2pm}
\mathcal{A}^{I, (\pm)} &=& \tilde{G}^I{}_{i j}\Gamma^{i j} \mp 2 L^I + 2i\tilde{\nabla}_i X^I \Gamma^i - 6i\chi \bigg(Q^{I J} - \frac{2}{3}X^{I}X^{J}\bigg)V_J \ .
\eea
These are derived from the naive restriction of the supercovariant derivative and the algebraic KSE on ${\cal S}$.
Furthermore, if $\eta_{-}$ solves $(\ref{covr})$ then
\bea
\eta_+ = \Gamma_{+}\Theta_{-}\eta_{-}~,
\label{epfem}
\eea
also solves $(\ref{covr})$. However, further analysis using global techniques, is required in order to determine if $\Theta_-$
has a non-trivial kernel.

\newsection{Global Analysis: Lichnerowicz Theorems}
\label{maxpex}

In this section, we shall establish a correspondence between parallel spinors $\eta_\pm$ satisfying
({\ref{covr}}), and spinors in the kernel of appropriately defined horizon Dirac operators.
We define the horizon Dirac operators associated with the supercovariant derivatives following from the gravitino KSE as
\bea
{\mathscr D}^{(\pm)} \equiv \Gamma^{i}\nabla_{i}^{(\pm)} = \Gamma^{i}\tilde{\nabla}_{i} + \Psi^{(\pm)}~,
\eea
where
\bea
\label{alg3pm}
\Psi^{(\pm)} \equiv \Gamma^{i}\Psi^{(\pm)}_{i} = \mp \frac{1}{4}h_i\Gamma^{i}  \mp \frac{3i}{4}\alpha  - \frac{3i}{8}\tilde{F}_{\ell_1 \ell_2}\Gamma^{\ell_1 \ell_2}
- \frac{3i}{2}\chi V_{I}\tilde{A}^{I}{}_{i}\Gamma^{i} + \frac{3}{2}\chi V_I X^{I} \ .
\eea

To establish the Lichnerowicz type theorems, we begin by calculating the Laplacian of $\parallel \eta_\pm \parallel^2$. Here we will assume throughout that  ${\mathscr D}^{(\pm)}\eta_\pm=0$, so
\bea
\label{lapla}
\tilde{\nabla}^i \tilde{\nabla}_i ||\eta_{\pm}||^2 = 2{\rm Re } \langle\eta_\pm,\tilde{\nabla}^i \tilde{\nabla}_i\eta_\pm\rangle + 2 {\rm Re } \langle\tilde{\nabla}^i \eta_\pm, \tilde{\nabla}_i \eta_\pm\rangle \ .
\eea
To evaluate this expression note that
\bea
\tilde{\nabla}^i \tilde{\nabla}_i \eta_\pm &=& \Gamma^{i}\tilde{\nabla}_{i}(\Gamma^{j}\tilde{\nabla}_j \eta_\pm) -\Gamma^{i j}\tilde{\nabla}_i \tilde{\nabla}_j \eta_\pm
\nonumber \\
&=& \Gamma^{i}\tilde{\nabla}_{i}(\Gamma^{j}\tilde{\nabla}_j \eta_\pm) + \frac{1}{4}\tilde{R}\eta_\pm
\nonumber \\
&=& \Gamma^{i}\tilde{\nabla}_{i}(-\Psi^{(\pm)}\eta_\pm) + \frac{1}{4}\tilde{R} \eta_\pm \ .
\eea
Therefore the first term in (\ref{lapla}) can be written as,
\bea
\label{lap1}
{\rm Re } \langle\eta_\pm,\tilde{\nabla}^i \tilde{\nabla}_i\eta_\pm \rangle = \frac{1}{4}\tilde{R}\parallel \eta_\pm \parallel^2
+ {\rm Re } \langle\eta_\pm, \Gamma^{i}\tilde{\nabla}_i(-\Psi^{(\pm)})\eta_\pm\rangle
+ {\rm Re } \langle\eta_\pm, \Gamma^{i}(-\Psi^{(\pm)})\tilde{\nabla}_i \eta_\pm \rangle~.
\nonumber \\
\eea
For the second term in (\ref{lapla}) we write,
\bea
\label{lap2}
\hspace{-1cm}{\rm Re } \langle\tilde{\nabla}^i \eta_\pm, \tilde{\nabla}_i \eta_\pm\rangle 
&=& \parallel {\nabla^{(\pm)}}\eta_{\pm} \parallel^2 - 2{\rm Re } \langle \eta_{\pm}, \Psi^{(\pm)i\dagger}\tilde{\nabla}_{i}\eta_{\pm}\rangle - {\rm Re } \langle \eta_\pm , \Psi^{(\pm)i\dagger}\Psi^{(\pm)}_i \eta_\pm \rangle  .
\eea
We remark that  $\dagger$ is the adjoint with respect to the $Spin_{c}(3)$-invariant inner product ${\rm Re } \langle \phantom{i},\phantom{i} \rangle$.\footnote{This inner product is positive definite and symmetric.}
Therefore using (\ref{lap1}) and (\ref{lap2}) with (\ref{lapla}) we have,
\bea
\label{extralap1b}
\frac{1}{2}\tilde{\nabla}^i \tilde{\nabla}_i ||\eta_{\pm}||^2 &=& \parallel {\nabla^{(\pm)}}\eta_{\pm} \parallel^2 + {\rm Re } \langle \eta_\pm, \bigg(\frac{1}{4}\tilde{R} + \Gamma^{i}\tilde{\nabla}_i(-\Psi^{(\pm)}) - \Psi^{(\pm)i\dagger}\Psi^{(\pm)}_i \bigg) \eta_\pm \rangle
\nonumber \\
&+& {\rm Re } \langle \eta_\pm, \bigg( \Gamma^{i}(-\Psi^{(\pm)}) - 2\Psi^{(\pm)i\dagger}\bigg)\tilde{\nabla}_i \eta_\pm \rangle \ .
\eea
In order to simplify the expression for the Laplacian, we observe that the second line in ({\ref{extralap1b}}) can be rewritten as
\bea
\label{bilin}
{\rm Re } \langle \eta_\pm, \bigg( \Gamma^{i}(-\Psi^{(\pm)}) - 2\Psi^{(\pm)i\dagger}\bigg)\tilde{\nabla}_i \eta_\pm \rangle = {\rm Re } \langle \eta_\pm, \mathcal{F}^{(\pm)}\Gamma^{i}\tilde{\nabla}_i \eta_\pm \rangle \pm\frac{1}{2}h^i \tilde{\nabla}_i \parallel \eta_\pm \parallel^2~,
\eea
where 
\bea
\mathcal{F}^{(\pm)} = \mp\frac{1}{4}h_{j}\Gamma^{j} \pm \frac{i}{4}\alpha + \frac{i}{8}\tilde{F}_{\ell_1 \ell_2}\Gamma^{\ell_1 \ell_2}
- \frac{3i}{2} \chi V_{I}\tilde{A}^{I}{}_{\ell}\Gamma^{\ell}
- \frac{5}{2}\chi V_{I}X^{I} \ .
\eea
 We also have the following identities 
\bea
\label{hermiden1}
{\rm Re } \langle \eta_+, \Gamma^{\ell_1 \ell_2} \eta_+ \rangle = {\rm Re } \langle \eta_+, \Gamma^{\ell_1 \ell_2 \ell_3} \eta_+ \rangle = 0
\eea
and
\bea
\label{hermiden2}
{\rm Re } \langle \eta_+, i\Gamma^{\ell} \eta_+ \rangle = 0 \  .
\eea
It follows that
\bea
\label{laplacian}
\frac{1}{2}\tilde{\nabla}^i \tilde{\nabla}_i ||\eta_{\pm}||^2 &=& \parallel {{\nabla}^{(\pm)}}\eta_{\pm} \parallel^2 \pm\frac{1}{2}h^i \tilde{\nabla}_{i}\parallel \eta_\pm \parallel^2
\nonumber \\
&+& {\rm Re } \langle \eta_\pm, \bigg(\frac{1}{4}\tilde{R} + \Gamma^{i}\tilde{\nabla}_i(-\Psi^{(\pm)}) - \Psi^{(\pm)i\dagger}\Psi^{(\pm)}_i  + \mathcal{F}^{(\pm)}(-\Psi^{(\pm)})\bigg) \eta_\pm \rangle \ .
\nonumber \\
\eea
It is also useful to evaluate ${\tilde{R}}$ using (\ref{feq4}); we obtain
\bea
\tilde{R} &=& -\tilde{\nabla}^{i}(h_i) + \frac{1}{2}h^2 + \frac{3}{2}\alpha^2
+ \frac{3}{4}\tilde{F}^2 - 2\chi^2 U 
\nonumber \\
&+& Q_{I J}\bigg(\tilde{\nabla}^{i}{X^I} \tilde{\nabla}_{i}{X^J} + L^{I}L^{J} + \frac{1}{2}\tilde{G}^{I}{}_{\ell_1 \ell_2}\tilde{G}^{J \ell_1 \ell_2}\bigg) \ .
\eea
One obtains, upon using the field equations and Bianchi identities,
\bea
\label{quad}
& &\bigg(\frac{1}{4}\tilde{R} + \Gamma^{i}\tilde{\nabla}_i(-\Psi^{(\pm)})  - \Psi^{(\pm)i\dagger}\Psi^{(\pm)}_i  + \mathcal{F}^{(\pm)}(-\Psi^{(\pm)}) \bigg)\eta_\pm
\nonumber \\
&=& \bigg[ \frac{3i}{2} \chi V_{I}\tilde{\nabla}^{\ell}{(\tilde{A}^{I}{}_{\ell})}
\mp \frac{3i}{4} \chi V_{I}\tilde{A}^{I}{}_{\ell}h^{\ell} \mp \frac{9i}{4} \chi V_{I}X^{I} \alpha + \big(\pm \frac{1}{4}\tilde{\nabla}_{\ell_1}(h_{\ell_2}) \mp \frac{3}{16}\alpha \tilde{F}_{\ell_1 \ell_2}\big)\Gamma^{\ell_2 \ell_2} 
\nonumber \\
&+& i\big(\pm \frac{3}{4}\tilde{\nabla}_{\ell}(\alpha) + \frac{3}{4}\tilde{\nabla}^{j}(\tilde{F}_{j \ell}) - \frac{1}{8}h_{\ell}\alpha \mp \frac{1}{4}h^{j}\tilde{F}_{j \ell} - \frac{3}{2} \chi^2 V_{J}X^{J}V_{I}\tilde{A}^{I}{}_{\ell}\big)\Gamma^{\ell} 
\nonumber \\
&+& \frac{3}{8} \chi V_{I}\tilde{A}^{I}{}_{\ell_1}\tilde{F}_{\ell_2 \ell_3}\Gamma^{\ell_1 \ell_2 \ell_3}\bigg]\eta_{\pm}
\nonumber \\
&+&  \bigg(\frac{1}{8}Q_{I J}\tilde{G}^{I \ell_1 \ell_2}\tilde{G}^{J}{}_{\ell_1 \ell_2}
+ \frac{1}{4}Q_{I J}L^{I}L^{J} 
+ \frac{9}{4}\chi^2  V_I V_J Q^{I J} 
- \frac{3}{2}\chi^2 V_I V_J X^{I}X^{J}
\nonumber \\
&+& \frac{1}{4}Q_{I J}\tilde{\nabla}_{\ell}{X^{I}}\tilde{\nabla}^{\ell}{X^{J}}
+ \frac{3i}{8}\tilde{G}^{I}{}_{\ell_1 \ell_2}\tilde{\nabla}_{\ell_3}{X_{I}}\Gamma^{\ell_1 \ell_2 \ell_3}
-\frac{3}{2} \chi  V_{I} \tilde{\nabla}_{\ell}X^{I}\Gamma^{\ell}
+ \frac{3i}{4} \chi V_{I} \tilde{G}^{I}{}_{\ell_1 \ell_2}\Gamma^{\ell_1 \ell_2}
\bigg)\eta_\pm
\nonumber \\
&-& {1 \over 4} \big(1 \mp 1\big) {\tilde{\nabla}}^i (h_i)  \eta_\pm  \ .
\eea
One can show that the fourth and fifth line in (\ref{quad}) can be written in terms of the algebraic KSE (\ref{alg2pm}), in particular we find,
\bea
\label{blah}
\frac{1}{16}Q_{I J}{\cal A}^{I, (\pm)\dagger}{\cal A}^{J, (\pm)}\eta_\pm &=& \bigg(\frac{1}{8}Q_{I J}\tilde{G}^{I \ell_1 \ell_2}\tilde{G}^{J}{}_{\ell_1 \ell_2}
+ \frac{1}{4}Q_{I J}L^{I}L^{J} + \frac{9}{4}\chi^2  V_I V_J Q^{I J} 
\nonumber \\
&-& \frac{3}{2}\chi^2 V_I V_J X^{I}X^{J} + \frac{1}{4}Q_{I J}\tilde{\nabla}_{\ell}{X^{I}}\tilde{\nabla}^{\ell}{X^{J}} 
+ \frac{3i}{8}\tilde{G}^{I}{}_{\ell_1 \ell_2}\tilde{\nabla}_{\ell_3}{X_{I}}\Gamma^{\ell_1 \ell_2 \ell_3}
\nonumber \\
&-&\frac{3}{2} \chi  V_{I} \tilde{\nabla}_{\ell}X^{I}\Gamma^{\ell}
+ \frac{3i}{4} \chi V_{I} \tilde{G}^{I}{}_{\ell_1 \ell_2}\Gamma^{\ell_1 \ell_2}
\bigg)\eta_\pm \ .
\eea
Note that on using (\ref{hermiden1}) and (\ref{hermiden2}) all the terms on the RHS of the above expression, with the exception of the final three lines, vanish in the second line of (\ref{laplacian}) since all these terms in ({\ref{quad}}) are anti-Hermitian.
Also, for $\eta_+$ the final line in ({\ref{quad}}) also vanishes and thus there is no contribution to the Laplacian of $\parallel \eta_+ \parallel^2$ in (\ref{laplacian}). For $\eta_{-}$ the final line in ({\ref{quad}}) does give an extra term in the Laplacian of $\parallel \eta_- \parallel^2$ in (\ref{laplacian}). For this reason, the analysis of the conditions imposed by the global properties of ${\cal{S}}$ is different in these two cases and thus we will consider the Laplacians of $\parallel \eta_\pm \parallel^2$ separately.

For the Laplacian of $\parallel \eta_+ \parallel^2$, we obtain from ({\ref{laplacian}}):
\bea
\label{l1}
{\tilde{\nabla}}^{i}{\tilde{\nabla}}_{i}\parallel\eta_+\parallel^2 - \, h^i {\tilde{\nabla}}_{i}\parallel\eta_+\parallel^2 = 2\parallel{{\nabla}^{(+)}}\eta_{+}\parallel^2 +~  \frac{1}{16}Q_{I J} {\rm Re } \langle {\cal A}^{I, (+)} \eta_+, {\cal A}^{J, (+)} \eta_+ \rangle  \ .
\eea
The maximum principle thus implies that $\eta_+$ are Killing spinors on ${\cal{S}}$ assuming that it is compact, connected and without boundary, i.e. 
\bea
{{\nabla}^{(+)}}\eta_{+}=0, \quad {\cal A}^{I, (+)}\eta_{+} = 0
\eea
and moreover $\parallel\eta_+\parallel=\mathrm{const}$. 

The Laplacian of $\parallel \eta_- \parallel^2$
is calculated from ({\ref{laplacian}}), on taking account of the contribution to the second line of
({\ref{laplacian}}) from the final line of ({\ref{quad}}). One
obtains
\bea
\label{l2}
{\tilde{\nabla}}^{i} \bigg(\tilde{\nabla}_{i} \parallel \eta_- \parallel^2 + \parallel \eta_- \parallel^2  h_{i}\bigg)
= 2 \parallel{{\nabla}^{(-)}}\eta_{-}\parallel^2 +~ \frac{1}{16}Q_{I J} {\rm Re } \langle {\cal A}^{I,(-)} \eta_-, {\cal A}^{J,(-)}\eta_- \rangle ~.
\eea
On integrating this over ${\cal{S}}$ and assuming that ${\cal{S}}$ is compact and without boundary, the LHS vanishes since it is a total derivative and one finds that $\eta_{-}$ are Killing spinors on ${\cal{S}}$, i.e 
\bea
{{\nabla}^{(-)}}\eta_{-}=0, \qquad {\cal A}^{I, (-)}\eta_{-} = 0 \ .
\eea

This establishes the Lichnerowicz type theorems for both positive and negative chirality spinors $\eta_\pm$ which are in the kernels of the horizon Dirac operators ${{\mathscr D}}^{(\pm)}$: i.e.

\bea
\{ \ {{\nabla}^{(\pm)}}\eta_{\pm}=0, \quad {\rm and} \quad {\cal A}^{I, (\pm)}\eta_{\pm} = 0 \ \}
\qquad \Longleftrightarrow \qquad {{\mathscr D}}^{(\pm)} \eta_\pm = 0 \ .
\eea

\newsection{Supersymmetry Enhancement}

In this section we will consider the counting of the number of supersymmetries, which will differ slightly in the ungauged and gauged case.  We will denote by $N_\pm$ the number
of linearly independent (over $\bC$) $\eta_\pm$ Killing spinors i.e,
\bea
N_\pm={{\rm dim}}_{\mathbb{C}}  \  {\rm{Ker}} \{ {{\nabla}^{(\pm)}}, {\cal A}^{I, (\pm)} \} ~.
\eea

Consider a spinor $\eta_+$ satisfying the corresponding KSEs in ({\ref{covr}}).
In the ungauged theory, the spinor $C*\eta_+$ also satisfies the same KSEs, and $C* \eta_+$ is linearly independent from $\eta_+$, where $C*$ denotes charge conjugation. So in the ungauged theory, $N_+$ must be even. However, in the gauged theory $C*\eta_+$ is not parallel, so $N_+$ need not be even.

The spinors in the KSEs of $N=2, D=5$ (un)gauged supergravity horizons with an arbitrary number of vector multiplets are Dirac spinors. In terms of the spinors $\eta_\pm$ restricted to ${\cal{S}}$, for the ungauged theory the spin bundle $\bS$ decomposes
as $\bS = \bS^+ \oplus \bS^-$ where the signs refer to the projections with respect to $\Gamma_\pm$,
and $\bS^\pm$ are $Spin(3)$ bundles.
For the gauged theory, the spin bundle $\bS \otimes {\cal L}$, where ${\cal L}$ is a $U(1)$ bundle on ${\cal{S}}$, decomposes as $\bS \otimes {\cal L} = \bS^+ \otimes {\cal L} \oplus \bS^- \otimes {\cal L}$ where $\bS^\pm \otimes {\cal L}$ are $Spin_{c}(3) = Spin(3) . U(1)$.

To proceed further, we will show that the analysis which we have developed implies that the
number of real supersymmetries of near-horizon geometries is $4 N_+$. This is because the number
of real supersymmetries is $N=2 (N_+ + N_-)$ and we shall establish that $N_+ = N_-$ via the following global analysis. In particular, utilizing the Lichnerowicz type theorems which we have established previously, we have
\bea
N_\pm=\mathrm{dim}\,\mathrm{Ker}\, {\mathscr D}^{(\pm)}~.
\eea

Next let us focus on the index of the ${\mathscr D}^{(+)}$ operator. Since ${\mathscr D}^{(+)}$  is defined on the odd dimensional manifold ${\cal S}$, the index vanishes \cite{atiyah1}.  As a result, we conclude that
\bea
\mathrm{dim}\,\mathrm{Ker}\, {\mathscr D}^{(+)}= \mathrm{dim}\,\mathrm{Ker}\, ({\mathscr D}^{(+)})^\dagger
\eea
where $({\mathscr D}^{(+)})^\dagger$ is the adjoint of ${\mathscr D}^{(+)}$.  Furthermore observe that
\bea
\Gamma_- ({\mathscr D}^{(+)})^\dagger= {\mathscr D}^{(-)} \Gamma_-~,
\eea
and so
\bea
N_-=\mathrm{dim}\,\mathrm{Ker}\, ({\mathscr D}^{(-)})=\mathrm{dim}\,\mathrm{Ker}\, ({\mathscr D}^{(+)})^\dagger~.
\eea
Therefore, we conclude that $N_+=N_-$ and so the number of (real) supersymmetries of such horizons is $N=2 (N_++N_-) = 4 N_+$.

\subsection{Algebraic Relationship between $\eta_+$ and $\eta_{-}$ Spinors}
\label{kernal}

We shall exhibit the existence of the $\mathfrak{sl}(2,\bR)$ symmetry of gauged $D=5$ vector multiplet horizons by directly constructing
the vector fields on the spacetime which generate the action of $\mathfrak{sl}(2,\bR)$. The existence of these vector fields
is a direct consequence of the doubling of the supersymmetries. We have seen that if $\eta_-$ is a Killing spinor, then  $\eta_+=\Gamma_+\Theta_-\eta_-$
is also a Killing spinor provided that $\eta_+\not=0$. It turns out that under certain conditions this is always possible. To consider this we must investigate the kernel
of $\Theta_-$.

\vskip 0.3cm
{\it Lemma:} Suppose that ${\cal S}$ and the fields satisfy the requirements for the maximum principle to apply,
and that 
\bea
\mathrm{Ker}\, \Theta_- \neq \{0\}~.
\label{kerz}
\eea
\vskip 0.3cm
Then the near-horizon data is trivial, i.e. all fluxes vanish and the scalars are constant.
\vskip2mm
{\it Proof:}
Suppose that there is $\eta_-\not=0$
such that $\Theta_- \eta_-=0$. In such a case,
  ({\ref{int3}}) gives
$ \Delta {\rm Re } \langle \eta_- , \eta_- \rangle =0$. Thus $\Delta =0$,
as $\eta_-$ is no-where vanishing. Next, the gravitino KSE $\nabla^{(-)}\eta_-=0$, together with ${\rm Re } \langle \eta_{-}, \Gamma_{i}\Theta_{-}\eta_{-} \rangle=0$, imply that
\begin{eqnarray}
\label{nrm1a}
{\tilde{\nabla}}_i \parallel \eta_-\parallel^2 = - h_i  \parallel \eta_-\parallel^2~.
\end{eqnarray}
On taking the divergence of this expression, eliminating  ${\tilde{\nabla}}^i h_i$ upon using ({\ref{feq3}}),
and after setting $\Delta=0$, one finds
\begin{eqnarray}
\label{nrm1ab}
{\tilde{\nabla}}^i {\tilde{\nabla}}_i  \parallel \eta_-\parallel^2 &=&
 \bigg(2\alpha^2 + \frac{1}{2}\tilde{F}^2 + \frac{4}{3}Q_{I J}L^{I}L^{J} + \frac{1}{3}Q_{I J}\tilde{G}^{I \ell_1 \ell_2}\tilde{G}^{J}{}_{\ell_1 \ell_2}
 + \frac{4}{3}\chi^2 U \bigg)  \parallel \eta_-\parallel^2~.
\nonumber \\
\end{eqnarray}
As we have assumed that $Q_{IJ}$ is positive definite, and that $U \geq 0$, the maximum principle implies that $\parallel \eta_- \parallel^2$ is constant. We conclude
that $\alpha = \tilde{F} = L^{I} = \tilde{G}^{I} = U = 0$ and from ({\ref{int7}}) that $X^I$ is constant. Also $U = 0$ implies $V_{I} = 0$. Furthermore, ({\ref{nrm1a}}) implies that $dh=0$, and then (\ref{auxeq1}) implies that $\beta = M^{I} = 0$.
Finally, integrating ({\ref{feq3}}) over the horizon section implies that $h=0$. Thus, all the fluxes vanish, and the scalars are constant. {\rightline{ $\square$}}

We remark that in the ungauged theory, if $\mathrm{Ker}\, \Theta_- \neq \{0\}$, triviality
of the near-horizon data implies that the spacetime geometry is
${\mathbb{R}}^{1,1} \times T^3$. In the case of the gauged theory, imposing $\mathrm{Ker}\, \Theta_- \neq \{0\}$ leads directly to a contradiction. To see this, note that the condition $U=0$ implies
that
\bea
V_I V_J (X^I X^J-{1 \over 2} Q^{IJ})=0 \ .
\eea
However the algebraic KSE imply that
\bea
V_I V_J (Q^{IJ}-{2 \over 3} X^I X^J)=0 \ .
\eea
These conditions cannot hold simultaneously, so there is a contradiction.

Hence, to exclude both the trivial ${\mathbb{R}}^{1,1} \times T^3$
solution in the ungauged theory, and the contradiction in the gauged theory, we shall henceforth take $\mathrm{Ker}\, \Theta_-=\{0\}$.

\subsection{The $\mathfrak{sl}(2,\bR)$ Symmetry}

Having established how to obtain $\eta_+$ type spinors from $\eta_-$ spinors, we next proceed
to determine the $\mathfrak{sl}(2,\bR)$ spacetime symmetry.
First note that the spacetime Killing spinor $\epsilon$ can be expressed in terms of $\eta_\pm$ as
\begin{eqnarray}
\epsilon= \eta_++ u \Gamma_+\Theta_-\eta_-+ \eta_-+r \Gamma_-\Theta_+\eta_++ru \Gamma_-\Theta_+\Gamma_+\Theta_-\eta_-~.
\label{gensolkse}
\end{eqnarray}
Since the $\eta_-$ and $\eta_+$ Killing spinors appear in pairs for supersymmetric horizons, let us choose a $\eta_-$ Killing spinor.  Then from the previous results, horizons with non-trivial fluxes also admit $\eta_+=\Gamma_+\Theta_-\eta_-$ as a Killing spinor. Taking $\eta_-$ and $\eta_+=\Gamma_+\Theta_-\eta_-$,
one can construct two linearly independent Killing spinors on the  spacetime as
\bea
\epsilon_1=\eta_-+u\eta_++ru \Gamma_-\Theta_+\eta_+~,~~~\epsilon_2=\eta_++r\Gamma_-\Theta_+\eta_+~.
\eea
It is known from the general theory of supersymmetric $D=5$ backgrounds that for any Killing spinors $\zeta_1$ and $\zeta_2$ the dual vector field $K(\zeta_1, \zeta_2)$ of the 1-form
bilinear
\bea
\omega(\zeta_1, \zeta_2) &=& {\rm Re } \langle(\Gamma_+-\Gamma_-) \zeta_1, \Gamma_a\zeta_2\rangle\, e^a 
\label{1formbi}
\eea
is a Killing vector which leaves invariant all the other bosonic fields of the theory, i.e.
\bea
{\cal L}_{K} g = {\cal L}_{K} X^{I} = {\cal L}_{K} F^{I} = 0 \ .
\eea
Evaluating the 1-form bilinears of the Killing spinor $\epsilon_1$ and $\epsilon_2$, we find that
\begin{eqnarray}
\label{bforms}
\omega_1(\epsilon_1, \epsilon_2)&=& (2r {\rm Re} \langle\Gamma_+\eta_-, \Theta_+\eta_+\rangle+  4 u r^2  \parallel \Theta_{+}\eta_+\parallel^2) \,{\bf{e}}^+-2u \parallel\eta_+\parallel^2\, {\bf{e}}^-
\cr
&+& ({\rm Re} \langle \Gamma_+\eta_{-}, \Gamma_{i}\eta_{+}\rangle + 4 u r {\rm Re} \langle \eta_+, \Gamma_{i} \Theta_+ \eta_+ \rangle) {\bf{e}}^i~,
 \cr
 \omega_2(\epsilon_2, \epsilon_2)&=& 4 r^2 \parallel \Theta_+ \eta_+\parallel^2 \,{\bf{e}}^+-2 \parallel\eta_+\parallel^2 {\bf{e}}^- + 4 r{\rm Re}  \langle \eta_{+}, \Gamma_{i} \Theta_+ \eta_+ \rangle {\bf{e}}^i~,
 \cr
 \omega_3(\epsilon_1, \epsilon_1)&=&(2\parallel\eta_-\parallel^2+4r u {\rm Re} \langle\Gamma_+\eta_-, \Theta_+\eta_+\rangle+ 4 r^2 u^2 \parallel \Theta_+ \eta_+\parallel^2 ) {\bf{e}}^+
 \cr
 &-& 2u^2 \parallel\eta_+\parallel^2 {\bf{e}}^-+(2u {\rm Re } \langle \Gamma_+ \eta_- , \Gamma_i \eta_+ \rangle
+ 4 u^2 r{\rm Re} \langle \eta_+, \Gamma_i \Theta_+ \eta_+ \rangle) {\bf{e}}^i~.
\nonumber \\
\end{eqnarray}
Moreover, we can establish the following identities
\begin{eqnarray}
\label{ident1}
- \Delta\, \parallel\eta_+\parallel^2 +4  \parallel\Theta_+ \eta_+\parallel^2 =0~,~~~{\rm Re } \langle \eta_+ , \Gamma_i \Theta_+ \eta_+ \rangle  =0~,
\end{eqnarray}
which follow from the first integrability condition in ({\ref{int1}}),  $\parallel\eta_+\parallel=\mathrm{const}$ and the KSEs of $\eta_+$.
Further simplification to the bilinears can be obtained by making use of (\ref{ident1}).
We then obtain
\begin{eqnarray}
 \omega_1(\epsilon_1, \epsilon_2)&=& (2r {\rm Re} \langle\Gamma_+\eta_-, \Theta_+\eta_+\rangle+  u r^2 \Delta \parallel \eta_+\parallel^2) \,{\bf{e}}^+-2u \parallel\eta_+\parallel^2\, {\bf{e}}^-+ \tilde V_i {\bf{e}}^i~,
 \cr
 \omega_2(\epsilon_2, \epsilon_2)&=& r^2 \Delta\parallel\eta_+\parallel^2 \,{\bf{e}}^+-2 \parallel\eta_+\parallel^2 {\bf{e}}^-~,
 \cr
 \omega_3(\epsilon_1, \epsilon_1)&=&(2\parallel\eta_-\parallel^2+4r u {\rm Re} \langle\Gamma_+\eta_-, \Theta_+\eta_+\rangle+ r^2 u^2 \Delta \parallel\eta_+\parallel^2) {\bf{e}}^+
 \cr
 && \qquad\qquad \qquad\qquad -2u^2 \parallel\eta_+\parallel^2 {\bf{e}}^-+2u \tilde V_i {\bf{e}}^i~,
 \label{b1forms}
\end{eqnarray}
where we have set
\begin{eqnarray}
\label{vii}
\tilde V_i =  {\rm Re } \langle \Gamma_+ \eta_- , \Gamma_i \eta_+ \rangle\, ~.
\end{eqnarray}

To uncover explicitly the $\mathfrak{sl}(2,\mathbb{R})$ symmetry of such horizons it remains to compute the Lie bracket algebra of the vector fields $K_1$, $K_2$ and $K_3$ which are dual to  the 1-form spinor bilinears $\omega_1, \omega_2$ and $\omega_3$. In simplifying the resulting expressions, we shall make use of the following identities
\begin{eqnarray}
&&-2 \parallel\eta_+\parallel^2-h_i \tilde V^i+2 {\rm Re } \langle\Gamma_+\eta_-, \Theta_+\eta_+\rangle=0~,~~~i_{\tilde V} (dh)+2 d {\rm Re } \langle\Gamma_+\eta_-, \Theta_+\eta_+\rangle=0~,
\cr
&& 2 {\rm Re } \langle\Gamma_+\eta_-, \Theta_+\eta_+\rangle-\Delta \parallel\eta_-\parallel^2=0~,~~~
{\tilde V}+ \parallel\eta_-\parallel^2 h+d \parallel\eta_-\parallel^2=0~.
\label{conconx}
\end{eqnarray}

We then obtain the following dual Killing vector fields:
\begin{eqnarray}
K_1 &=&-2u \parallel\eta_+\parallel^2 \partial_u+ 2r \parallel\eta_+\parallel^2 \partial_r+ \tilde V~,
\cr
K_2 &=&-2 \parallel\eta_+\parallel^2 \partial_u~,
\cr
K_3 &=&-2u^2 \parallel\eta_+\parallel^2 \partial_u +(2 \parallel\eta_-\parallel^2+ 4ru \parallel\eta_+\parallel^2)\partial_r+ 2u \tilde V~.
\label{kkk}
\end{eqnarray}

As we have previously mentioned, each of these Killing vectors also leaves invariant all the other bosonic fields in the theory. It is then straightforward to determine the
algebra satisfied by these isometries:

\vskip5mm

{\it Theorem:} The Lie bracket algebra of  $K_1$, $K_2$ and $K_3$  is $\mathfrak{sl}(2,\bR)$.
\vskip 0.3cm
{\it Proof:} Using the identities summarised above, one can demonstrate after a direct computation that
\begin{eqnarray}
[K_1,K_2]=2 \parallel\eta_+\parallel^2 K_2~,~~~[K_2, K_3]=-4 \parallel\eta_+\parallel^2 K_1  ~,~~~[K_3,K_1]=2 \parallel\eta_+\parallel^2 K_3~. \ \
\nonumber \\
\end{eqnarray}

\subsection{Isometries of ${\cal S}$}

It is known that the vector fields associated with the 1-form Killing spinor bilinears given in (\ref{1formbi}) leave invariant all the fields of
gauged $D=5$ supergravity with vector multiplets.  In particular suppose that $\tilde V \neq 0$. The isometries $K_a$ ($a=1,2,3$) leave all the bosonic fields invariant:
\bea
{\cal L}_{K_a} g=0, \qquad {\cal L}_{K_a} F^I=0, \qquad {\cal L}_{K_a} X^I=0 \ .
\eea
Imposing these conditions and expanding in $u,r$, and also making use of the identities
({\ref{conconx}}), one finds that
\begin{eqnarray}
\tilde\nabla_{(i} \tilde V_{j)}=0~,~~~ {\cal L}_{\tilde V} h= {\cal L}_{\tilde V}\Delta=0~,~~~ {\cal L}_{\tilde V} X^{I} = 0~, \quad
 {\cal L}_{\tilde V} \tilde{F}= {\cal L}_{\tilde V} \alpha= {\cal L}_{\tilde V} L^{I}= {\cal L}_{\tilde V} \tilde{G}^{I}=0~.
\nonumber \\
\end{eqnarray}
Therefore $\tilde V$ is an isometry of ${\cal S}$ and leaves all the fluxes on ${\cal S}$ invariant. In fact,${\tilde{V}}$ is a spacetime
isometry as well. Furthermore, the conditions ({\ref{conconx}}) imply that ${\cal L}_{\tilde V}\parallel\eta_-\parallel^2=0$. 

\subsection{Solutions with $\tilde V=0$}

A special case arises for $\tilde V=0$, where the group action generated by $K_1, K_2$ and $K_3$ has only 2-dimensional orbits. A direct substitution of this condition in (\ref{conconx}) reveals that
\bea
\label{zva}
\Delta \parallel\eta_-\parallel^2=2 \parallel\eta_+\parallel^2~,~~~h=\Delta^{-1} d\Delta~.
\eea
Since $h$ is exact, such horizons are static. A coordinate transformation $r\rightarrow \Delta r$ reveals that the  geometry is a warped product of $AdS_2$ with ${\cal S}$, $AdS_2\times_w {\cal S}$.

To further investigate these solutions, in particular in the gauged theory,
it will be useful to define the 1-form spinor bilinear $Z$ on ${\cal{S}}$ by
\bea
Z_i = \langle \eta_+, \Gamma_i \eta_+ \rangle
\eea
We remark that as a consequence of Fierz identities, this bilinear satisfies
\bea
Z^2 = (\parallel \eta_+ \parallel^2)^2
\eea
and in what follows we shall without loss of generality set $\parallel \eta_+ \parallel=1$. Furthermore, ({\ref{zva}}) implies that $\Delta$ is positive everywhere on ${\cal{S}}$. To proceed note that ({\ref{ident1}}) implies 
\bea
h- {\tilde{\star}} {\tilde{F}} = 2 \chi V_I X^I Z
\eea
where ${\tilde{\star}}$ denotes the Hodge dual on ${\cal{S}}$. This condition can be used to eliminate ${\tilde{F}}$ from the reduced gravitino KSE on ${\cal{S}}$,
({\ref{covr}}), and one obtains the condition
\bea
{\tilde{\nabla}}^i \big( \Delta^{-2} Z_i \big) = -6 \Delta^{-2} \chi V_I X^I
\eea
on setting $Z^2=1$, and using ({\ref{zva}}) to eliminate $h$ in terms of $d \Delta$.
Integrating this expression over ${\cal{S}}$ gives
\bea
\int_{{\cal{S}}} \Delta^{-2} \chi V_I X^I =0
\eea
So, for the case of the gauged theory, there must exist a point on ${\cal{S}}$ at which $V_I X^I=0$. However, at such a point $U=-{9 \over 2} Q^{IJ} V_I V_J <0$, in contradiction to our assumption that $U \geq 0$ on ${\cal{S}}$. Hence, it follows that there are no near-horizon geometries in the gauged theory
for which ${\tilde{V}}=0$.

\section{Conclusion}

We have investigated the supersymmetry preserved by horizons in $N=2, D=5$ gauged, and ungauged, supergravity with an arbitrary number of vector multiplets. Making use of global techniques, we have demonstrated that such horizons always admit $N=4N_+$ (real) supersymmetries. Furthermore, in the ungauged theory,
we have shown that $N_+$ must be even. Therefore, all supersymmetric near-horizon geometries in the ungauged theory must be maximally supersymmetric. We have also shown that the near-horizon geometries possess a $\mathfrak{sl}{(2, \mathbb{R})}$ symmetry group. The analysis that we have conducted is further evidence that this type of symmetry enhancement is a generic property of supersymmetric black holes.

In fact, the complete classification of the geometries in the ungauged theory is quite straightforward, because the identity
\bea
K_2=-2 \parallel\eta_+\parallel^2 \partial_u~.
\eea
implies that the timelike isometry $\partial_u$ can be written as a spinor bilinear. All supersymmetric near-horizon geometries in the ungauged theory for which $\partial_u$
can be written as a spinor bilinear in this fashion have been fully classified in
\cite{gutbh}. In particular, the solutions
reduce to those of the minimal ungauged theory and the scalars are constant. 
The supersymmetry enhancement in this case therefore automatically imposes  an attractor-type mechanism, whereby the scalars take constant values on the horizon.

The possible near-horizon geometries
in the ungauged theory are therefore ${\mathbb{R}}^{1,1} \times T^3$; and $AdS_3 \times S^2$, 
corresponding to the near-horizon black string/ring geometry \cite{Chamseddine:1999qs, BS1, BR1}; and the near-horizon BMPV solution \cite{BMPV1, BMPV2}.
For near-horizon solutions in the gauged theory, the total number of supersymmetries is either 4 or 8. In the case of maximal supersymmetry,
the geometry is locally isometric to $AdS_5$, with $F^I=0$ and constant scalars.{\footnote{As observed in \cite{preons}, there also exist discrete quotients of $AdS_5$ preserving 6 out of 8 supersymmetries. In this case, the spinors which are excluded are not smooth due to the periodic identification.}} 

It remains to 
classify the geometries of $N=4$ solutions in the gauged theory; details of this will be given elsewhere. We have shown that the horizon sections of these solutions
admit at least one rotational isometry ${\tilde{V}}$, which is a symmetry of
the full solution. It would be interesting to determine if additional isometries
also exist. This is because the analysis in \cite{Kunduri:2007qy} provides a complete classification of near-horizon geometries of supersymmetric black holes of $U(1)^3$-gauged supergravity with vector multiplets, assuming the existence of two commuting rotational isometries on the horizon section. In this case, the classification for the geometry of the horizon shows that it is either spherical $S^3$, $S^1 \times S^2$ or a $T^3$ - the last two have no analogue in the minimal gauged theory, corresponding to the near-horizon geometry $AdS_3 \times S^2$ and $AdS_3 \times T^2$. The difference between
the minimal theory and the $STU$ theory in this context is encoded in the parameter
\bea
\lambda = Q^{IJ} V_I V_J - (V_I X^I)^2
\eea

The near-horizon geometries constructed in \cite{Kunduri:2007qy} for which
$S^1 \times S^2$ arises as a solution are required to have $\lambda>0$ as
a consequence of the analysis of the geometry. This condition can be satisfied
in the $STU$ theory, but not in minimal gauged supergravity. In fact,
supersymmetric $AdS_5$ black rings have been excluded from minimal gauged supergravity in \cite{5dindex}. This analysis did not assume the existence of two commuting
rotational isometries, rather it derived the existence of such isometries via
the supersymmetry enhancement mechanism. The possibility of 
an $AdS_5$ black ring remains for the gauged $STU$ theory. 
As we have noted,
a regular supersymmetric near-horizon geometry with $S^1 \times S^2$ event horizon
topology is known to exist in the gauged $STU$ theory. There are no known 
obstructions, analogous to the stability analysis considered in \cite{tdef},
to extending the near-horizon solution into the bulk, and it is unknown if
a supersymmetric $AdS_5$ black ring exists.

Another avenue for further research is higher derivative supergravity. In general, higher derivative supergravity theories have extremely complicated field equations, which makes
a systematic analysis of the near-horizon geometries challenging. One theory for which the
field equations are relatively simple is heterotic supergravity with $\alpha'$ corrections, 
the near-horizon analysis in this theory has already been considered in \cite{Fontanella}. 
In the context of $D=5$ theories, higher derivative theories have been constructed in \cite{Hanaki},
and the near-horizon analysis has been considered in \cite{Gutowskihd}, however the analysis 
in this case assumes that the black hole timelike isometry ${\partial \over \partial u}$
arises as a Killing spinor bilinear. The analysis of the KSEs is relatively straightforward, because the gravitino equation has the same form as in the 2-derivative theory. However,
the 2-form which appears in the gravitino equation is an auxiliary field which is related to the
Maxwell field strengths via highly nonlinear auxiliary field equations. This makes the
analysis of the geometric conditions particularly involved. Despite these difficulties,
it would nevertheless be interesting to investigate supersymmetry enhancement of near-horizon geometries
in higher derivative supergravity.

%
%

\vskip 0.5cm
\noindent{\bf Acknowledgements} \vskip 0.1cm
\noindent
UK is supported by a STFC PhD fellowship.
\vskip 0.5cm

\setcounter{section}{0}\setcounter{equation}{0}
\appendix{Supersymmetry Conventions}

We first present a matrix representation of $\mathrm{Cliff}(4,1)$ adapted to the basis ({\ref{basis1}}).
The space of Dirac spinors is identified with $\bC^4$ and we set
\bea
\Gamma_i = \begin{pmatrix}  \sigma^i \ \ \ \ \  0  \\   \ \ 0 \ \ -\sigma^i \end{pmatrix}, \qquad
\Gamma_- = \begin{pmatrix} \ \ 0  \ \ \ \sqrt{2}\, \bI_2  \\  0 \ \ \ \ \ 0 \end{pmatrix}, \qquad
\Gamma_+ = \begin{pmatrix}  \ \ 0 \ \ \ \ \ 0  \\ \sqrt{2}\, \bI_2 \ \ \ \ 0 \end{pmatrix}
\eea
where $\sigma^i$, $i=1,2,3$ are the Hermitian Pauli matrices $\sigma^i \sigma^j = \delta^{ij} \bI_2 + i \epsilon^{ijk} \sigma^k$.
Note that
\bea
\Gamma_{+-} = \begin{pmatrix} -\bI_2 \ \ \ \ \ 0  \\ \ \ \ 0 \ \ \ \bI_2 \end{pmatrix}~,
\eea
and hence
\bea
\Gamma_{+-123} = -i \bI_4~.
\eea
It will be convenient to decompose the spinors into positive and negative chiralities
with respect to the lightcone directions as
\bea
\epsilon = \epsilon_+ + \epsilon_-~,
\eea
where
\bea
\Gamma_{+-} \epsilon_\pm = \pm \epsilon_\pm \ , \qquad {\rm or \ equivalently} \qquad \Gamma_\pm \epsilon_\pm =0~.
\eea
With these conventions, note that
\bea
\label{conv}
\Gamma_{ij} \epsilon_\pm = \mp i \epsilon_{ij}{}^k \Gamma_k \epsilon_\pm \ ,
\qquad \Gamma_{ijk} \epsilon_\pm = \mp i \epsilon_{ijk} \epsilon_\pm~.
\eea

The Dirac representation of $Spin(4,1)$ decomposes under $Spin(3)=SU(2)$ as $\bC^4=\bC^2\oplus \bC^2$ each subspace
specified by the lightcone projections $\Gamma_\pm$. On each $\bC^2$,
we have made use of the $Spin(3)$-invariant inner product  ${\rm Re } \langle , \rangle$ which is identified with the standard Hermitian
inner product. On $\bC^2\oplus \bC^2$, the Lie algebra of
$Spin(3)$ is spanned  by $\Gamma_{ij}$, $i,j=1,2,3$. In particular,  note that $(\Gamma_{ij})^\dagger = - \Gamma_{ij}$.

The charge conjugation operator $C$ 
can be chosen to be 
\bea
C=\begin{pmatrix}  i \sigma^2 \ \ \ \ \  0  \\   \ \ 0 \ \ -i \sigma^2 \end{pmatrix} = i \Gamma_{2}
\eea
and satisfies $C* \Gamma_\mu + \Gamma_\mu C* =0$. Furthermore, if $\epsilon$ is any Dirac spinor then
\bea
\langle \epsilon, C* \epsilon \rangle =0 \ .
\eea

\appendix{Spin Connection and Curvature}

The non-vanishing components of the spin connection in the frame basis ({\ref{basis1}}) are
\begin{eqnarray}
&&\Omega_{-,+i} = -{1 \over 2} h_i~,~~~
\Omega_{+,+-} = -r \Delta, \quad \Omega_{+,+i} ={1 \over 2} r^2(  \Delta h_i - \partial_i \Delta),
\cr
&&\Omega_{+,-i} = -{1 \over 2} h_i, \quad \Omega_{+,ij} = -{1 \over 2} r dh_{ij}~,~~~
\Omega_{i,+-} = {1 \over 2} h_i, \quad \Omega_{i,+j} = -{1 \over 2} r dh_{ij},
\cr
&&\Omega_{i,jk}= \tilde\Omega_{i,jk}~,
\end{eqnarray}
where $\tilde\Omega$ denotes the spin-connection of the 3-manifold ${{\cal{S}}}$ with basis ${\bf{e}}^i$.
If $f$ is any function of spacetime, then frame derivatives are expressed in terms of co-ordinate derivatives  as
\begin{eqnarray}
\label{frco}
\partial_+ f &=& \partial_u f +{1 \over 2} r^2 \Delta \partial_r f~,~~
\partial_- f = \partial_r f~,~~
\partial_i f = {\tilde{\partial}}_i f -r \partial_r f h_i \ .
\end{eqnarray}
The non-vanishing components of the Ricci tensor in the
 basis ({\ref{basis1}}) are
\bea
R_{+-} &=& {1 \over 2} \tn^i h_i - \Delta -{1 \over 2} h^2~,~~~
R_{ij} = {\tilde{R}}_{ij} + \tn_{(i} h_{j)} -{1 \over 2} h_i h_j
\nonumber \\
R_{++} &=& r^2 \big( {1 \over 2} \tn^2 \Delta -{3 \over 2} h^i \tn_i \Delta -{1 \over 2} \Delta \tn^i h_i + \Delta h^2
+{1 \over 4} (dh)_{ij} (dh)^{ij} \big)
\nonumber \\
R_{+i} &=& r \big( {1 \over 2} \tn^j (dh)_{ij} - (dh)_{ij} h^j - \tn_i \Delta + \Delta h_i \big) \ ,
\eea
where $\tn$ denotes the Levi-Civita connection of ${\cal S}$, and  ${\tilde{R}}$ is the Ricci tensor of the horizon section ${\cal S}$, and $i,j$ denote $\bbe^i$ frame indices.

\appendix{Horizon Bianchi Identities and Field Equations}
Substituting the fields (\ref{nhf}) into the the Bianchi identity $dF^I = 0$ implies
\bea
\label{beqo}
\beta^I = (d_h \alpha^I), ~~d\tilde{F}^I = 0
\eea
and
\bea
\label{beqo2}
d\beta^{I} + \alpha^{I}dh + d\alpha^{I} \wedge h = 0 \ .
\eea
Note that (\ref{beqo2}) is implied (\ref{beqo}). Similarly, the independent field equations of the near horizon fields are as follows. The Maxwell gauge equations (\ref{maxwell}) are given by,
\bea
d_h(Q_{I J}\star_3 \tilde{F}^J) - Q_{I J}\star_3\beta^J =  \frac{1}{2}C_{I J K}\alpha^J \tilde{F}^K \ .
\eea
In components this can be expressed as,
\bea
\label{feq1o}
\tilde{\nabla}^j(Q_{I J}\tilde{F}^J{}_{j i}) - Q_{I J}h^j \tilde{F}^J{}_{j i} + Q_{I J}\beta^J{}_i + \frac{1}{4}C_{I J K}\epsilon_i{}^{\ell_1 \ell_2}\alpha^J \tilde{F}^K{}_{\ell_1 \ell_2} = 0
\eea
which corresponds to the $i$-component of (\ref{maxeq}). There is another equation given by the $+$-component of (\ref{maxeq}) but this is implied by (\ref{feq1o}) and is not used in the analysis at any stage.
The $+-$ and $ij$-component of the Einstein equation (\ref{eins}) gives
\bea
\label{feq2o}
- \Delta - \frac{1}{2}h^2 + \frac{1}{2}\tilde{\nabla}^i(h_i) = -Q_{I J}\bigg(\frac{2}{3}\alpha^I \alpha^J + \frac{1}{6}\tilde{F}^I{}_{\ell_1 \ell_2}\tilde{F}^{J \ell_1 \ell_2}\bigg) - \frac{2}{3}\chi^2 U
\eea
and
\bea
\label{feq3o}
\tilde{R}_{i j} &=& -\tilde{\nabla}_{(i}h_{j)} + \frac{1}{2}h_i h_j -\frac{2}{3}\chi^2 U \delta_{i j} 
\nonumber \\
&+& Q_{I J}\bigg[\tilde{F}^I{}_{i \ell}\tilde{F}^{J}{}_{j}{}^{\ell} + \tilde{\nabla}_i X^I \tilde{\nabla}_j X^J 
+ \delta_{i j}\bigg(\frac{1}{3}\alpha^I\alpha^J - \frac{1}{6}\tilde{F}^I{}_{\ell_1 \ell_2}\tilde{F}^{J \ell_1 \ell_2}\bigg) \bigg] \ .
\nonumber \\
\eea
The scalar field equation (\ref{scalareq1}) gives
\bea
\label{feq4o}
&&\hspace*{-0.5cm}\tilde{\nabla}^{i}\tilde{\nabla}_{i}{X_{I}} - h^i\tilde{\nabla}_i X_I + \tilde{\nabla}_{i}{X^{M}} \tilde{\nabla}^{i}{X^{N}} \left(\frac{1}{2}C_{M N K}X_{I} X^{K} - \frac{1}{6}C_{I M N}\right) 
\nonumber \\
&&\hspace{-1cm}+\hspace{0.5cm} \bigg[\frac{1}{2}\tilde{F}^{M}{}_{\ell_1 \ell_2} \tilde{F}^{N \ell_1 \ell_2} -\alpha^M \alpha^N \bigg] \bigg(C_{I N P} X_{M} X^{P} - \frac{1}{6}C_{I M N}-6X_{I} X_{M} X_{N}
+\frac{1}{6}C_{M N J} X_{I} X^{J}\bigg)  
\nonumber \\
&&\hspace{-1cm}+\hspace{0.5cm} 3 \chi^2 V_{M} V_{N}\bigg(\frac{1}{2}C_{I J K}Q^{M J}Q^{N K} + X_{I}(Q^{M N} - 2 X^{M}X^{N})\bigg) = 0 \ .
\eea
We remark that the $++$ and $+i$ components of the
Einstein equations, which are
\bea
\label{auxeq1o}
{1 \over 2} \tn^i \tn_i \Delta -{3 \over 2} h^i \tn_i \Delta-{1 \over 2} \Delta \tn^i h_i
+ \Delta h^2 +{1 \over 4} dh_{ij} dh^{ij} - Q_{I J}\beta^{I}{}_{\ell}\beta^{J \ell} = 0~,
\eea
and
\bea
\label{auxeq2o}
{1 \over 2} \tn^j dh_{ij}-dh_{ij} h^j - \tn_i \Delta + \Delta h_i
+ Q_{I J}\alpha^{I}\beta^{J}{}_{i} - Q_{I J}\beta^{I}{}_{\ell}\tilde{F}^{J}{}_{i}{}^{\ell} = 0
\eea
are implied by ({\ref{feq2o}}), ({\ref{feq3o}}), ({\ref{feq4o}}),
together with ({\ref{feq1o}}).
and the Bianchi identities ({\ref{beqo}}).

\appendix{Gauge Field Decomposition}
Using the decomposition $F^I = FX^I + G^I$ with $F = X_I F^I$, $X_I G^I = 0$ and $dF^I = 0$ implies
\bea
dF &=& -X_I dG^I
\nonumber \\
(\delta^I{}_J - X^I X_J) dG^J &=& -dX^I \wedge F \ .
\eea
We write the near-horizon fields as
\bea
F^I &=& \bbe^+ \wedge \bbe^- \alpha^I + r \bbe^+ \wedge \beta^I + {\tilde F}^I
\nonumber \\
F &=& \bbe^+ \wedge \bbe^- \alpha + r \bbe^+ \wedge \beta + {\tilde F}
\nonumber \\
G^I &=& \bbe^+ \wedge \bbe^- L^I + r \bbe^+ \wedge M^I + {\tilde G}^I \ ,
\eea
where $X_I L^I = X_I M^I = X_I \tilde{G}^I = 0$ and $\alpha = X_I \alpha^I$, $\tilde{F} = X_I \tilde{F}^I$, $\beta = X_I \beta^I$.
\bea
\label{decomp}
\alpha^I &=& \alpha X^I + L^I
\nonumber \\
\beta^I &=& \beta X^I + M^I
\nonumber \\
{\tilde F}^I &=& {\tilde F} X^I + {\tilde G}^I \ .
\eea
By using (\ref{decomp}) we can express the Bianchi identities (\ref{beqo}) as
\bea
\label{beq}
\beta &=& d_h \alpha - L^I dX_I 
\nonumber \\
d\tilde{F} &=& -X_I d\tilde{G}^I
\nonumber \\
(\delta^I{}_J - X^I X_J)(d_h L^{J} -M^{J}) &=& -dX^{I}\alpha
\nonumber \\
(\delta^I{}_J - X^I X_J) d\tilde{G}^J &=& -dX^I \wedge \tilde{F}
\eea
and corresponding to (\ref{beqo2})
\bea
\label{beq1}
d M^{I} - h \wedge M^{I} + L^{I} dh  + dX^{I} \wedge \beta &=& 0
\nonumber \\
d \beta - h \wedge \beta + \alpha  dh  + dX_{I} \wedge M^{I} &=& 0 \ .
\eea
However, (\ref{beq1}) is implied by (\ref{beq}). The field equations can also be decomposed using (\ref{decomp}) as follows. The Maxwell gauge equation (\ref{feq1o}) gives
\bea
\label{feq1}
&&\frac{3}{2}X_I \tilde{\nabla}^j(\tilde{F}_{j i}) + \tilde{\nabla}^j(Q_{I J}\tilde{G}^J{}_{j i}) + \frac{3}{2}\tilde{\nabla}^jX_I \tilde{F}_{j i} -\frac{3}{2}X_I h^j \tilde{F}_{j i} - Q_{I J}h^j \tilde{G}^J{}_{j i}
+ \frac{3}{2}X_I \beta_i 
\nonumber \\
&+& Q_{I J}M^J{}_i + \frac{1}{4}\epsilon_i{}^{\ell_1 \ell_2}\bigg(6 X_I \alpha \tilde{F}_{\ell_1 \ell_2} - 2 Q_{I J}\alpha \tilde{G}^J{}_{\ell_1 \ell_2} - 2 Q_{I J}\tilde{F}_{\ell_1 \ell_2}L^J + C_{I J K}L^J \tilde{G}^K{}_{\ell_1 \ell_2}\bigg) = 0
\nonumber \\
\eea
where we have used the identity $\tilde{\nabla}_i(Q_{I J})X^J = 3\tilde{\nabla}_i X_I$. By contracting with $X^I$ this gives,
\bea
\label{feq2}
\tilde{\nabla}^j(\tilde{F}_{j i}) + \tilde{\nabla}^j(X_J) \tilde{G}^J{}_{j i} - h^j \tilde{F}_{j i} + \beta_i 
+ \epsilon_i{}^{\ell_1 \ell_2} \alpha \tilde{F}_{\ell_1 \ell_2} - \frac{1}{3}Q_{I J} \epsilon_i{}^{\ell_1 \ell_2}L^I\tilde{G}^J{}_{\ell_1 \ell_2} = 0 \ .
\eea
The Einstein equation (\ref{feq2o}) gives
\bea
\label{feq3}
- \Delta - \frac{1}{2}h^2 + \frac{1}{2}\tilde{\nabla}^i(h_i) &=& -\bigg[\alpha^2 + \frac{1}{4}\tilde{F}_{\ell_1 \ell_2}\tilde{F}^{\ell_1 \ell_2} + \frac{2}{3}\chi^2 U 
\nonumber \\
&+& Q_{I J}\bigg(\frac{2}{3}L^I L^J + \frac{1}{6}\tilde{G}^I{}_{\ell_1 \ell_2}\tilde{G}^{J \ell_1 \ell_2}\bigg)\bigg]
\eea
and (\ref{feq3o})
\bea
\label{feq4}
\tilde{R}_{i j} &=& -\tilde{\nabla}_{(i}h_{j)} + \frac{1}{2}h_i h_j + \frac{3}{2}\tilde{F}_{i k}\tilde{F}_j{}^k + \delta_{i j}\bigg(\frac{1}{2}\alpha^2 - \frac{1}{4}\tilde{F}_{\ell_1 \ell_2}\tilde{F}^{\ell_1 \ell_2} -\frac{2}{3}\chi^2 U \bigg)
\nonumber \\
&+& Q_{I J}\bigg[\tilde{G}^{I}{}_{i \ell}\tilde{G}^{J}{}_{j}{}^{\ell} + \tilde{\nabla}_{i}{X^I} \tilde{\nabla}_{j}{X^J} + \delta_{i j}\bigg(\frac{1}{3}L^I L^J - \frac{1}{6}\tilde{G}^I{}_{\ell_1 \ell_2}\tilde{G}^{J \ell_1 \ell_2}\bigg)\bigg] \ .
\eea
The scalar field equations (\ref{feq4o}) give
\bea
\label{feq5}
&&\tilde{\nabla}^{i}{\tilde{\nabla}_{i}{X_{I}}} - h^i\tilde{\nabla}_i X_I + \tilde{\nabla}_{i}{X^{M}}\tilde{\nabla}^{i}{X^{N}} \bigg(\frac{1}{2}C_{M N K}X_{I}X^{K} - \frac{1}{6}C_{M N I}\bigg) 
\nonumber \\
&+& \frac{2}{3}Q_{I J}\bigg(2\alpha L^{J} - \tilde{F}_{\ell_1 \ell_2}\tilde{G}^{J \ell_1 \ell_2}\bigg)
 -\frac{1}{12}\bigg[\tilde{G}^{M}{}_{\ell_1 \ell_2}\tilde{G}^{N \ell_1 \ell_2} - 2 L^{M}L^{N}\bigg]\bigg(C_{M N I} - X_{I}C_{M N J}X^{J}\bigg) 
\nonumber \\
&+& 3 \chi^2 V_{M} V_{N}\bigg(\frac{1}{2}C_{I J K}Q^{M J}Q^{N K} + X_{I}(Q^{M N} - 2 X^{M}X^{N})\bigg) = 0
\eea
Furthermore (\ref{auxeq1o}) gives
\bea
\label{auxeq1}
{1 \over 2} \tn^i \tn_i \Delta -{3 \over 2} h^i \tn_i \Delta-{1 \over 2} \Delta \tn^i h_i
+ \Delta h^2 +{1 \over 4} dh_{ij} dh^{ij}  = \frac{3}{2}\beta^2 + Q_{I J}M^{I}{}_{\ell}M^{J \ell}~,
\eea
and ({\ref{auxeq2o}}) gives
\bea
\label{auxeq2}
{1 \over 2} \tn^j dh_{ij}-dh_{ij} h^j - \tn_i \Delta + \Delta h_i
= \frac{3}{2}\bigg(\beta_{\ell}\tilde{F}_{i}{}^{\ell}-\alpha \beta_{i} \bigg) + Q_{I J}\bigg(M^{I}{}_{\ell} \tilde{G}^{J}{}_{i}{}^{\ell} - L^{I}M^{J}{}_{i}\bigg) \ .
\nonumber \\
\eea
The conditions ({\ref{auxeq1}}) and ({\ref{auxeq2}}) correspond to the $++$ and $+i$-component of the Einstein equation and we remark that these are both implied by ({\ref{feq3}}), ({\ref{feq4}}), ({\ref{feq5}}), together with ({\ref{feq1}}) and ({\ref{feq2}}) and the Bianchi identities ({\ref{beq}}).

\appendix{Simplification of KSEs on ${\cal{S}}$}

In this appendix we show how several of the KSEs on ${\cal{S}}$ are implied
by the remaining KSEs, together with the field equations and Bianchi identities. To begin,
we show that
(\ref{int1}), (\ref{int2}), (\ref{int5}), and (\ref{int8}) which contain $\tau_+$ are implied from those containing $\phi_+$, along with some of the field equations and Bianchi identities. Then,
we establish that (\ref{int3}) and the terms linear in $u$ in (\ref{int4}) and (\ref{int7}) from the $+$ component are implied by the field equations, Bianchi identities and the $-$ component of (\ref{int4}) and (\ref{int7}). 

A particular useful identity is obtained by considering the integrability condition of (\ref{int4}), which implies that
\bea
(\tilde{\nabla}_{j}\tilde{\nabla}_{i} - \tilde{\nabla}_{i}\tilde{\nabla}_{j})\phi_{\pm} &=& \bigg(\pm\frac{1}{4}\tilde{\nabla}_j(h_i)  
\mp\frac{1}{4}\tilde{\nabla}_i(h_j)
\pm \frac{i}{4}\tilde{\nabla}_j(\alpha) \Gamma_i  
\mp \frac{i}{4}\tilde{\nabla}_i(\alpha) \Gamma_j
+ \frac{i}{2}\tilde{\nabla}_j(\tilde{F}_{i \ell})\Gamma^\ell
\nonumber \\
&-& \frac{i}{2}\tilde{\nabla}_i(\tilde{F}_{j \ell})\Gamma^\ell 
- \frac{i}{8}\tilde{\nabla}_j(\tilde{F}_{\ell_1 \ell_2})\Gamma_i{}^{\ell_1 \ell_2} 
+ \frac{i}{8}\tilde{\nabla}_i(\tilde{F}_{\ell_1 \ell_2})\Gamma_j{}^{\ell_1 \ell_2} 
\mp \alpha \tilde{F}_{j \ell}\Gamma_{i}{}^{\ell}
\nonumber \\
&\pm& \frac{1}{4}\alpha\tilde{F}_{i}{}^{\ell}\Gamma_{j}{}^{\ell}
+ \frac{1}{8}\tilde{F}_{j}{}^{\lambda}\tilde{F}_{\lambda \ell}\Gamma_{i}{}^{\ell}
- \frac{1}{8}\tilde{F}_{i}{}^{\lambda}\tilde{F}_{\lambda \ell}\Gamma_{j}{}^{\ell}
- \frac{3}{8}\tilde{F}_{i \ell_1}\tilde{F}_{j \ell_2}\Gamma^{\ell_1 \ell_2}
-\frac{1}{8}\alpha^2 \Gamma_{i j}
\nonumber \\
&+& \frac{1}{16}\tilde{F}^2 \Gamma_{i j}
+ \frac{1}{2}\chi^2 V_{I}V_{J}X^{I}X^{J} \Gamma_{i j}
\mp \frac{i}{2}\chi V_{I}X^{I} \alpha \Gamma_{i j}
- \chi V_{I} \Gamma_{[i}\tilde{\nabla}_{j]}{(X^{I})}
\nonumber \\
&-&  \frac{3i}{2} \chi V_{I}\tilde{F}^{I}{}_{i j}
+ i \chi V_{I}X^{I} \tilde{F}_{[i | \ell |}\Gamma_{j]}{}^{\ell}
\bigg)\phi_{\pm} \ \ .
\label{DDphicond}
\eea
This will be used in the analysis of (\ref{int1}), (\ref{int3}), (\ref{int5}) and the positive chirality part of  (\ref{int4}) which is linear in $u$. In order to show that the conditions are redundant, we will be considering different combinations of terms which vanish
as a consequence of the independent KSEs. However, non-trivial identities are found by 
explicitly expanding out the terms in each case.

\subsection{The  condition (\ref{int1})}
\label{int1sec}
It can be shown that the algebraic condition on $\tau_+$ (\ref{int1}) is implied by the independent KSEs. Let us define,
\bea
&&\xi_1 = \bigg({1\over2}\Delta - {1\over8}(dh)_{ij}\Gamma^{ij} -\frac{i}{4}\beta_i \Gamma^i + \frac{3i}{2}\chi V_{I}\alpha^I \bigg)\phi_+ 
\nonumber \\
&&+ 2\bigg({1\over4} h_i\Gamma^i - \frac{i}{8}(-\tilde{F}_{j k}\Gamma^{j k} + 4\alpha) + \frac{1}{2} \chi V_{I}X^{I}\bigg)\tau_+ \ ,
\eea
where $\xi_1=0$ is equal to the condition (\ref{int1}). 
It is then possible to show that this expression for $\xi_1$ can be re-expressed as
\bea
\label{int1cond}
&&\xi_1 = \bigg(-\frac{1}{4}\tilde{R} - \Gamma^{i j}\tilde{\nabla}_{i}\tilde{\nabla}_{j}\bigg)\phi_+ 
+ \mu_{I}\mathcal{A}^{I}{}_1 = 0
\eea
where the first two terms cancel as a consequence of the definition of curvature, and
\bea
\mu_{I} = \frac{3i}{16}\Gamma^{i}\tilde{\nabla}_{i}{X_{I}} - Q_{I J}\bigg(\frac{7}{24}L^{J} + \frac{5}{48}\tilde{G}^{J}{}_{\ell_1 \ell_2}\Gamma^{\ell_1 \ell_2} \bigg) + \frac{i}{8}\chi V_{I}
\eea
the scalar curvature is can be written as
\bea
\tilde{R} &=& -2\Delta - \frac{1}{2}h^{2} + \frac{7}{2}\alpha^2 + \frac{5}{4}\tilde{F}^2 - \frac{2}{3}\chi^2 U 
\nonumber \\
&+& Q_{I J}\bigg(\frac{7}{3}L^{I}L^{J} + \frac{5}{6}\tilde{G}^{I\ell_1 \ell_2}\tilde{G}^{J}{}_{\ell_1 \ell_2} + \tilde{\nabla}_{i}{X^{I}}\tilde{\nabla}^{i}{X^{J}}\bigg)
\eea
and
\bea
\label{int1condaux}
\mathcal{A}^{I}{}_1 = \bigg[\tilde{G}^I{}_{i j}\Gamma^{i j} - 2 L^I + 2i\tilde{\nabla}_i X^I \Gamma^i - 6i\chi \bigg(Q^{I J} - \frac{2}{3}X^{I}X^{J}\bigg)V_J\bigg]\phi_+ \ .
\eea
The expression appearing in ({\ref{int1condaux}}) vanishes because
 $\mathcal{A}^{I}{}_1 = 0$ is equivalent to the positive chirality part of (\ref{int7}).
Furthermore, the expression for $\xi_1$ given in ({\ref{int1cond}}) also vanishes.
We also use (\ref{DDphicond}) to evaluate the terms in the first bracket in (\ref{int1cond}) and explicitly expand out the terms with $\mathcal{A}^{I}{}_1$. In order to obtain (\ref{int1}) from these expressions we make use of the Bianchi identities (\ref{beq}), the field equations (\ref{feq1}) and (\ref{feq2}). We have also made use of the $+-$ component of the Einstein equation (\ref{feq4}) in order to rewrite the scalar curvature $\tilde{R}$ in terms of $\Delta$. Therefore (\ref{int1}) follows from (\ref{int4}) and (\ref{int7}) together with the field equations and Bianchi identities mentioned above.

\subsection{The  condition (\ref{int2})}
Here we will show that the algebraic condition on $\tau_+$ (\ref{int2}) follows from (\ref{int1}). It is convenient to define
\bea
\xi_2 = \bigg(\frac{1}{4}\Delta h_i \Gamma^{i} - \frac{1}{4}\partial_{i}\Delta \Gamma^{i}\bigg)\phi_+ + \bigg(-\frac{1}{8}(dh)_{ij}\Gamma^{ij} +\frac{3i}{4}\beta{}_i\Gamma^i  + \frac{3i}{2}\chi V_{I}\alpha^I\bigg) \tau_+ \ ,
\eea
where $\xi_2=0$ equals the condition (\ref{int2}).
One can show after a computation that this expression for $\xi_2$ can be re-expressed as
\bea
\xi_2 = -\frac{1}{4}\Gamma^{i}\tilde{\nabla}_{i}{\xi_1} + \frac{7}{16}h_{j}\Gamma^{j}\xi_1 = 0 \ ,
\eea
which vanishes because $\xi_1 = 0$ is equivalent to the condition (\ref{int1}). In order to obtain this, we use the Dirac operator $\Gamma^{i}\tilde{\nabla}_{i}$ to act on (\ref{int1}) and apply the Bianchi identities (\ref{beq}) with the field equations (\ref{feq1}), (\ref{feq2}) and (\ref{feq5}) to eliminate the terms which contain derivatives of the fluxes, and we can also use (\ref{int1}) to rewrite the $dh$-terms in terms of $\Delta$. We then impose the algebraic conditions (\ref{int7}) and (\ref{int8}) to eliminate the $\tilde{\nabla}_i X^I$-terms, of which some of the remaining terms will vanish as a consequence of (\ref{int1}). We then obtain the condition (\ref{int2}) as required, therefore it follows from section \ref{int1sec} above that (\ref{int2}) is implied by (\ref{int4}) and (\ref{int7}) together with the field equations and Bianchi identities mentioned above.

\subsection{The  condition (\ref{int5})}
Here we will show the differential condition on $\tau_+$ (\ref{int5}) is not independent. Let us define
\bea
\lambda_i &=& \tilde \nabla_i \tau_{+} + \bigg( -\frac{3}{4}h_i - \frac{i}{4}\alpha\Gamma_i - \frac{i}{8}\tilde{F}_{j k}\Gamma_i{}^{j k} + \frac{i}{2}\tilde{F}_{i j}\Gamma^j - \frac{3i}{2}\chi V_{I}\tilde{A}^{I}{}_{i} - \frac{1}{2}\chi V_I X^{I}\Gamma_i\bigg )\tau_{+}
\nonumber \\
&&+ \bigg(-\frac{1}{4}(dh)_{ij}\Gamma^{j} - \frac{i}{4}\beta_j \Gamma_i{}^j + \frac{i}{2}\beta_i   \bigg)\phi_{+} \ , 
\eea
where $\lambda_i=0$ is equivalent to the condition (\ref{int5}). We can re-express this expression for
$\lambda_i$ as
\bea
\label{int5cond}
\lambda_ i = \bigg(-\frac{1}{4}\tilde{R}_{i j}\Gamma^{j} + \frac{1}{2}\Gamma^{j}(\tilde{\nabla}_{j}\tilde{\nabla}_{i} - \tilde{\nabla}_{i}\tilde{\nabla}_{j}) \bigg)\phi_+  + \frac{1}{2}\Lambda_{i, I}{\cal A}^{I}{}_{1} = 0~,
\eea
where the first terms again cancel from the definition of curvature, and
\bea
\Lambda_{i, I} = \frac{3i}{8} \tilde{\nabla}_{i}X_{I} +  Q_{I J}\bigg( \frac{1}{24}\tilde{G}^{J}{}_{\ell_1 \ell_2}\Gamma_{i}{}^{\ell_1 \ell_2} - \frac{1}{6}\tilde{G}^{J}{}_{i j}\Gamma^{j} - \frac{1}{12}L^{J}\Gamma_{i}\bigg) + \frac{i}{4}\chi V_{I} \Gamma_{i} \ ,
\eea
This vanishes as $\mathcal{A}^{I}{}_1 = 0$ is equivalent to the positive chirality  component of (\ref{int7}). The identity (\ref{int5cond}) is derived by making use of (\ref{DDphicond}), and explicitly expanding out the $\mathcal{A}^{I}{}_1$ terms. We can also evaluate (\ref{int5}) by substituting in (\ref{int6}) to eliminate $\tau_+$, and use (\ref{int4}) to evaluate the supercovariant derivative of $\phi_+$.  Then, on adding this to (\ref{int5cond}), one obtains a condition which vanishes identically on making use of the Einstein equation (\ref{feq4}). Therefore it follows that (\ref{int5}) is implied by the positive chirality component of (\ref{int4}), (\ref{int6}) and (\ref{int7}), the Bianchi identities (\ref{beq}) and the gauge field equations (\ref{feq1}) and (\ref{feq2}).

\subsection{The condition (\ref{int8})}
Here we will show that the algebraic condition containing $\tau_+$ (\ref{int8}) follows from the independent KSEs. We define 
\bea
\mathcal{A}^{I}{}_{2} &=& \bigg[\tilde{G}^I{}_{i j}\Gamma^{i j} + 2 L^I - 2i\tilde{\nabla}_i X^I \Gamma^i - 6i\chi \bigg(Q^{I J} - \frac{2}{3}X^{I}X^{J}\bigg)V_J \bigg]\tau_{+} + 2 M^I{}_i\Gamma^i \phi_{+}
\nonumber \\
\eea
and also set
\bea
\mathcal{A}_{I, 2} &=& Q_{I J}\mathcal{A}^{J}{}_{2} \ ,
\eea
where $\mathcal{A}^{I}{}_{2}=0$ equals the expression in (\ref{int8}). 
The expression for $\mathcal{A}_{I, 2}$ can be rewritten as
\bea
\label{nnaux1}
\mathcal{A}_{I, 2} &=& -\frac{1}{2}\Gamma^{i}\tilde{\nabla}_{i}{({\cal A}_{I, 1})}   
+ \Phi_{I J}{\cal A}^{J}{}_{1} 
\eea
where,
\bea
\Phi_{I J} &=& \bigg(-\frac{3}{4}Q_{J K}X_{I} - \frac{1}{8}C_{I J K}\bigg)\Gamma^{\ell}\tilde{\nabla}_{\ell}{X^{K}}
\nonumber \\
&+& \frac{i}{2}\bigg(\frac{1}{4}Q_{J K}X_{I} + \frac{1}{8}C_{I J K}\bigg)\bigg(\tilde{G}^{K}{}_{\ell_1 \ell_2}\Gamma^{\ell_1 \ell_2}
- 2 L^{K}\bigg)
\nonumber \\
&+& Q_{I J}\bigg(\frac{i}{16}\tilde{F}_{\ell_1 \ell_2}\Gamma^{\ell_1 \ell_2} - \frac{i}{8}\alpha + \frac{3}{8}h_{\ell}\Gamma^{\ell} + \frac{3i}{4}\chi V_{K}\tilde{A}^{K}{}_{\ell}\Gamma^{\ell} - \frac{3}{4}\chi V_{K}X^{K}\bigg) 
\nonumber \\
&+& \chi\bigg(-\frac{3}{8}C_{I J K}Q^{K M} - \frac{3}{4} X_{I}\delta^{M}{}_{J}\bigg)V_{M}~.
\eea
and $\mathcal{A}_{I, 1} = Q_{I J}\mathcal{A}^{J}{}_{1}$.
In evaluating the above conditions, we have made use of the $+$ component of (\ref{int4}) in order to evaluate the covariant derivative in the above expression. In addition we have made use of the Bianchi identities (\ref{beq}) and the field equations (\ref{feq1}), (\ref{feq2}) and (\ref{feq5}).

It follows from ({\ref{nnaux1}}) that $\mathcal{A}_{I, 2}=0$ as a consequence of the condition $\mathcal{A}_{I, 1}=0$, which as we have already noted is equivalent to the positive chirality part of ({\ref{int7}}).

\subsection{The  condition (\ref{int3})}
In order to show that (\ref{int3}) is implied by the independent KSEs, we define
\bea
\kappa &=& \bigg(-\frac{1}{2}\Delta - \frac{1}{8}(dh)_{ij}\Gamma^{ij} -\frac{3i}{4}\beta{}_i \Gamma^i + \frac{3i}{2}\chi V_{I}\alpha^I 
\nonumber \\
&+& 2\big(-{1\over4} h_i\Gamma^i - \frac{i}{8}(\tilde{F}_{j k}\Gamma^{j k} + 4\alpha) - \frac{1}{2}\chi V_{I}X^{I}\big) \Theta_{-} \bigg)\phi_{-} \ ,
\eea
where $\kappa$ equals the condition (\ref{int3}). Again, this expression can be rewritten as
\bea
\kappa = \bigg(\frac{1}{4}\tilde{R} + \Gamma^{i j}\tilde{\nabla}_{i}\tilde{\nabla}_{j}\bigg)\eta_{-} - \mu_{I}\mathcal{B}^{I}{}_1 = 0
\eea
where we use the (\ref{DDphicond}) to evaluate the terms in the first bracket, and
\bea
\mu_{I} = \frac{3i}{16}\Gamma^{i}\tilde{\nabla}_{i}{X_{I}} - Q_{I J}\bigg(-\frac{7}{24}L^{J} + \frac{5}{48}\tilde{G}^{J}{}_{\ell_1 \ell_2}\Gamma^{\ell_1 \ell_2} \bigg) + \frac{i}{8}\chi V_{I} \ .
\eea
 The expression above vanishes identically since the negative chirality component of (\ref{int7}) is equivalent to $\mathcal{B}^{I}{}_1 = 0$. In order to obtain (\ref{int3}) from these expressions we make use of the Bianchi identities (\ref{beq}) and the field equations (\ref{feq1}),(\ref{feq2}) and (\ref{feq5}). Therefore (\ref{int3}) follows from (\ref{int4}) and (\ref{int7}) together with the field equations and Bianchi identities mentioned above.

\subsection{The positive chirality part of (\ref{int4}) linear in $u$}
Since $\phi_+ = \eta_+ + u\Gamma_{+}\Theta_{-}\eta_-$, we must consider the part of the positive chirality component of (\ref{int4}) which is linear in $u$. We begin by defining
\bea
\mathcal{B}_{I, 1} &=& \bigg[\tilde{G}^I{}_{i j}\Gamma^{i j} + 2 L^I + 2i\tilde{\nabla}_i X^I \Gamma^i - 6i\chi \bigg(Q^{I J} - \frac{2}{3}X^{I}X^{J}\bigg)V_J\bigg]\eta_{-} \ .
\eea
We then determine that $\mathcal{B}_{I, 1}$ satisfies the following expression
\bea
\label{int4cond}
\bigg(\frac{1}{2}\Gamma^{j}(\tilde{\nabla}_{j}\tilde{\nabla}_{i} - \tilde{\nabla}_{i}\tilde{\nabla}_{j}) - \frac{1}{4}\tilde{R}_{i j}\Gamma^{j}  \bigg)\eta_{-} + \frac{1}{2}\Lambda_{i, I}{\cal B}^{I}{}_{1} = 0 ~,
\eea
where $\mathcal{B}_{I, 1} = Q_{I J}\mathcal{B}^{J}{}_{1}$, and
\bea
\Lambda_{i, I} = \frac{3i}{8} \tilde{\nabla}_{i}X_{I} +  Q_{I J}\bigg( \frac{1}{24}\tilde{G}^{J}{}_{\ell_1 \ell_2}\Gamma_{i}{}^{\ell_1 \ell_2} - \frac{1}{6}\tilde{G}^{J}{}_{i j}\Gamma^{j} + \frac{1}{12}L^{J}\Gamma_{i}\bigg) + \frac{i}{4}\chi V_{I} \Gamma_{i} \ .
\eea
We note that $\mathcal{B}_{I, 1}=0$ is equivalent to the negative chirality component of (\ref{int7}). 
Next, we use (\ref{DDphicond}) to evaluate the terms in the first bracket in (\ref{int4cond}) and explicitly expand out the terms with $\mathcal{B}^{I}{}_1$. The resulting expression corresponds to the expression obtained by expanding out the $u$-dependent part of the positive chirality component of (\ref{int4}) by using the 
negative chirality component of (\ref{int4}) to evaluate the covariant derivative. We have made use of the Bianchi identities (\ref{beq}) and the gauge field equations (\ref{feq1}) and (\ref{feq2}).

\subsection{The positive chirality part of condition (\ref{int7}) linear in $u$}
Again, as $\phi_+ = \eta_+ + u\Gamma_{+}\Theta_{-}\eta_-$, we must consider the part of the positive chirality component of (\ref{int7}) which is linear in $u$. 
One finds that the $u$-dependent part of (\ref{int7}) is proportional to
\bea
-\frac{1}{2}\Gamma^{i}\tilde{\nabla}_{i}{({\cal B}_{I, 1})}   
+ \Phi_{I J}{\cal B}^{J}{}_{1} \ ,
\eea
where,
\bea
\Phi_{I J} &=& \bigg(-\frac{3}{4}Q_{J K}X_{I} - \frac{1}{8}C_{I J K}\bigg)\Gamma^{\ell}\tilde{\nabla}_{\ell}{X^{K}}
\nonumber \\
&+& \frac{i}{2}\bigg(\frac{1}{4}Q_{J K}X_{I} + \frac{1}{8}C_{I J K}\bigg)\bigg(\tilde{G}^{K}{}_{\ell_1 \ell_2}\Gamma^{\ell_1 \ell_2}
+ 2 L^{K}\bigg)
\nonumber \\
&+& Q_{I J}\bigg(\frac{i}{16}\tilde{F}_{\ell_1 \ell_2}\Gamma^{\ell_1 \ell_2} + \frac{i}{8}\alpha + \frac{1}{8}h_{\ell}\Gamma^{\ell}+ \frac{3i}{4}\chi V_{K}\tilde{A}^{K}{}_{\ell}\Gamma^{\ell} - \frac{3}{4}\chi V_{K}X^{K}\bigg) 
\nonumber \\
&+& \chi\bigg(-\frac{3}{8}C_{I J K}Q^{K M} - \frac{3}{4} X_{I}\delta^{M}{}_{J}\bigg)V_{M}~.
\eea
and where we use the (\ref{DDphicond}) to evaluate the terms in the first bracket. In addition we have made use of the Bianchi identities (\ref{beq}) and the field equations (\ref{feq1}), (\ref{feq2}) and (\ref{feq5}).

\appendix{Scalar Orthogonality Condition}

In this appendix, we shall prove that if $L_{I}\partial_{a}{X^{I}} = 0$ for all 
values of $a=1, \dots, k-1$, i.e if $L_{I}$ is perpendicular to all $\partial_{a}{X^{I}}$, then it must be parallel to $X_{I}$. 

To establish the first result, it is sufficient to prove that the elements of
the set $\{ \partial_{a}{X^{I}} \, , a=1,\dots,k-1 \}$ are linearly independent.
Given this, the condition $L_{I}\partial_{a}{X^{I}} = 0$ for all 
values of $a=1, \dots, k-1$ implies that $L_I$ is orthogonal to all linearly independent
$k-1$ elements of this set, and hence must be parallel to the 1-dimensional orthogonal
complement to the set, which is parallel to $X_I$.

\vskip5mm
It remains to prove the following Lemma.
\vskip5mm

{\it Lemma:} The elements of the set $\{ \partial_{a}{X^{I}} \, , a=1,\dots,k-1 \}$ are linearly independent.

\vskip5mm

{\it Proof:} Let $N^{a}$ for $a=1,\dots,k-1$ be constants, where at least one is non-zero and suppose $N^{a}\partial_{a}{X^{I}} = 0$, then we have from (\ref{pbmetric})
\begin{eqnarray}
h_{a b}N^{a} = Q_{I J}\partial_{a}{X^{I}}\partial_{b}{X^{J}}N^{a} = 0
\end{eqnarray}
as $h_{a b}$ is non-degenerate, this implies that $N^{a} = 0$ for all $a=1,\dots,k-1$, which is a contradiction to our assumption that not all are zero and thus the elements of the set are linearly independent.

{\hskip150mm $\square$}

\vskip 0.5cm

We remark that an equivalent statement implied by the above reasoning is that 
if $L^{I}\partial_{a}{X_{I}} = 0$ for all $a=1, \dots k-1$ then $L^{I}$ must be parallel to $X^{I}$.

\end{document}